\documentclass[article,preprintnumbers,nofootinbib,showpacs]{revtex4}

\usepackage{epsfig}
\usepackage{amsmath}
\usepackage{verbatim}
\usepackage{psfrag}
\usepackage{bm}
\usepackage{bbm}
\usepackage{graphicx}
\usepackage{latexsym}
\usepackage[english]{babel}
\usepackage{amsthm}
\usepackage{color}
\usepackage{amssymb}

\begin{document}

\title{\Large Entropy and Correlators in Quantum Field Theory}

\preprint{ITP-UU-09/51, SPIN-09/42}

\

\preprint{HD-THEP-09-24}

\

\preprint{NORDITA-2009-69}


\author{Jurjen F. Koksma}
\email[]{J.F.Koksma@uu.nl, T.Prokopec@uu.nl,
M.G.Schmidt@thphys.uni-heidelberg.de} \affiliation{Institute for
Theoretical Physics (ITP) \& Spinoza Institute, Utrecht
University, Postbus 80195, 3508 TD Utrecht, The Netherlands}
\affiliation{NORDITA, Roslagstullsbacken 23, SE-106 91 Stockholm,
Sweden}

\author{Tomislav Prokopec}
\email[]{J.F.Koksma@uu.nl, T.Prokopec@uu.nl,
M.G.Schmidt@thphys.uni-heidelberg.de} \affiliation{Institute for
Theoretical Physics (ITP) \& Spinoza Institute, Utrecht
University, Postbus 80195, 3508 TD Utrecht, The Netherlands}
\affiliation{NORDITA, Roslagstullsbacken 23, SE-106 91 Stockholm,
Sweden}

\author{Michael G. Schmidt}
\email[]{J.F.Koksma@uu.nl, T.Prokopec@uu.nl,
M.G.Schmidt@thphys.uni-heidelberg.de} \affiliation{Institut f\"ur
Theoretische Physik, Heidelberg University, Philosophenweg 16,
D-69120 Heidelberg, Germany} \affiliation{NORDITA,
Roslagstullsbacken 23, SE-106 91 Stockholm, Sweden}

\begin{abstract}
It is well known that loss of information about a system, for some
observer, leads to an increase in entropy as perceived by this
observer. We use this to propose an alternative approach to
decoherence in quantum field theory in which the machinery of
renormalisation can systematically be implemented: neglecting
observationally inaccessible correlators will give rise to an
increase in entropy of the system. As an example we calculate the
entropy of a general Gaussian state and, assuming the observer's
ability to probe this information experimentally, we also
calculate the correction to the Gaussian entropy for two specific
non-Gaussian states.
\end{abstract}

\maketitle

\section{Introduction}
\label{Introduction}

The most natural way of defining the entropy of a quantum system
is to use the von Neumann entropy:
\begin{equation}
S_{\rm vN} = - {\rm Tr}[\hat \rho\ln(\hat \rho)]\,,
\label{entropy:vN}
\end{equation}
where $\hat\rho$ denotes the density operator which in the
Schr\"odinger picture satisfies the von Neumann equation:
\begin{equation}
\imath \hbar\frac{\partial}{\partial t}\hat \rho  = [\hat H,\hat
\rho] \,, \label{density op:eom}
\end{equation}
where $\hat H$ is the Hamiltonian. Since quantum mechanics and
quantum field theory are unitary theories, the von Neumann entropy
is conserved, albeit in general not zero.

The decoherence program~\cite{Zeh:1970,Joos:1984uk} usually
assumes the existence of some environment that is barely
observable to some observer. This allows us to construct a reduced
density operator $\hat\rho_{\mathrm{red}}$ characterising the
system S, which is obtained by tracing over the unobservable
environmental degrees of freedom E: $\hat\rho_{\mathrm{red}} ={\rm
Tr}_E[\hat \rho]$. This is a non-unitary process which
consequently generates entropy. This operation is justified by the
observer's inability to see the environmental degrees of freedom
(or better: access the information stored in the ES-correlators).
Yet, tracing out some degrees of freedom results in complications.
The simple looking von Neumann equation~(\ref{density op:eom}) is
replaced by an equation for the reduced density operator. This
``master equation''~\cite{PazZurek:2001,Zurek:2003zz} can be
solved for only in very simple situations, which has hampered
progress in decoherence studies in the context of interacting
quantum field theories. In particular, we are not aware of any
known solution of the master equation that would include
perturbative corrections and implement the program of
renormalisation. Also, one has no control of the error in a
calculation in the perturbative sense induced by neglecting
non-Gaussian corrections.

Here we propose a decoherence program that can be implemented in a
field theoretical setting and that also allows us to incorporate
renormalisation procedures. The idea is very simple, and we
present it for a real scalar field $\phi(x)$. A generalisation to
other types of fields, e.g. gauge and fermionic fields, should be
quite straightforward.

The density matrix $\hat{\rho}(t)$ contains all information about
the (possibly mixed) state a quantum system is in. From the
density matrix one can, in principle, calculate various
correlators. For example, the three Gaussian correlators are given
by:
\begin{subequations}
\label{3 equal time correlators}
\begin{eqnarray}
\langle \hat{\phi}(\vec{x}) \hat{\phi}(\vec{y}) \rangle &=& {\rm
Tr}[\hat \rho(t) \hat\phi(\vec x)\hat\phi(\vec y)] =
F(\vec{x},t;\vec{y},t')|_{t=t'}\label{3 equal time correlatorsa}
\\
\langle \hat{\pi}(\vec{x}) \hat{\pi}(\vec{y}) \rangle &=& {\rm
Tr}[\hat \rho(t) \hat\pi(\vec x)\hat\pi(\vec y)] =
\partial_{t} \partial_{t'} F(\vec{x},t;\vec{y},t')|_{t=t'} \label{3 equal time
correlatorsb}
\\
\frac{1}{2} \langle \{ \hat{\phi}(\vec{x}), \hat{\pi}(\vec{y}) \}
\rangle &=& \frac12 {\rm Tr}[\hat \rho(t)\{ \hat\phi(\vec
x),\hat\pi(\vec y)\}] = \partial_{t'}
F(\vec{x},t;\vec{y},t')|_{t=t'} \label{3 equal time
correlatorsc}\,,
\end{eqnarray}
\end{subequations}
where $\hat \pi$ denotes the momentum field conjugate to
$\hat\phi$, and where the curly brackets denote the
anti-commutator as usual. Note that these three Gaussian
correlators can all be determined from the statistical propagator
$F(\vec x,t;\vec y,t')$ in the Heisenberg picture as usual in
quantum field theory. The fourth Gaussian correlator is trivial as
it is restricted by the canonical commutation relations. There is
an infinite number of higher order, non-Gaussian correlators that
characterise a system, where one can think of e.g.:
\begin{subequations}
\label{NGCorrelators}
\begin{eqnarray}
\langle \hat{\phi}(\vec{x}_{1}) \cdots \hat{\phi}(\vec{x}_{n})
\rangle &=& {\rm Tr}[\hat \rho(t) \hat\phi(\vec{x}_{1}) \cdots
\hat\phi(\vec{x}_{n})] \label{NGCorrelatorsa}
\\
\langle \hat{\pi}(\vec{x}_{1}) \cdots \hat{\pi}(\vec{x}_{n})
\rangle &=& {\rm Tr}[\hat \rho(t) \hat\pi(\vec{x}_{1}) \cdots
\hat\pi(\vec{x}_{n})] \label{NGCorrelatorsb} \,,
\end{eqnarray}
where $n \geq 3$. Moreover, one can also imagine a higher order
correlation function consisting of a mixture of both
$\hat{\phi}$'s and $\hat{\pi}$'s. Finally, one could think of
other non-Gaussian correlators obtained by further differentiating
the statistical propagator:
\begin{equation}
\label{NGCorrelatorsc}
\partial_{t}^{n}\partial_{t'}^{m} F(\vec{x},t;\vec{y},t')|_{t=t'}
= \partial_{t}^n \partial_{t'}^{m} {\rm Tr}[\hat \rho(t_{0})
\hat\phi(\vec x,t)\hat\phi(\vec y,t')] |_{t=t'}\,,
\end{equation}
\end{subequations}
where $n+m \geq 3$. For free theories, all the non-Gaussian
correlators either vanish or can be expressed in terms of the
Gaussian correlators.

Let us turn our attention to interacting field theories and
present two simple arguments why, generally speaking, the Gaussian
correlators dominate over the non-Gaussian correlators. In
interacting quantum field theories, the Gaussian correlators all
stem from tree-level physics, whereas non-Gaussian correlators are
generated by the interactions. Relying on a perturbative treatment
of these interactions, we can safely expect that all of these
higher order, non-Gaussian correlators above are suppressed as
they generically are proportional to a non-zero power of the
(perturbatively small) interaction coefficient. A second, more
heuristic, argument invokes the (classical) central limit theorem,
according to which systems with many stochastic mutually
independent degrees of freedom tend to become normally
distributed, and hence nearly Gaussian. While this theorem is
derived for classical systems and for independent stochastic
variables, we believe that it will in a certain quantum disguise
also apply to weakly interacting quantum systems with many degrees
of freedom.

Hence, we expect that to a good approximation, many of the
relevant properties of quantum systems are encoded in the Gaussian
part $\hat \rho_{\rm{g}}$ of the density operator $\hat \rho$ and
the non-Gaussian correlators are thus typically much more
difficult to access\footnote{As a simple example, consider the
temperature correlations induced by scalar field perturbations
from inflation: while the amplitude of scalar field fluctuations
is given by ${\rm Tr}[\hat \rho(t) \hat\phi_N(\vec x,t)^2]\sim
G_NH^2\sim 10^{-12}$, where $\phi_N$ denotes the Newtonian
potential, where $G_N$ is Newton's constant and $H$ is the Hubble
rate during inflation, its quantum corrections are expected to be
of the order $\sim (G_NH^2)^2\sim 10^{-24}$ or smaller.}. The
crucial point is that all information about a general Gaussian
density matrix and thus about the Gaussian part of a state, is
stored only in the three equal time correlators (\ref{3 equal time
correlators}). All non-Gaussian correlators like
(\ref{NGCorrelators}) contain information about the correlations
between the system and environment as well as the environment's
correlations induced by higher order interactions. When projected
on the field amplitude basis in the Schr\"odinger picture, the
Gaussian part of a density matrix reads:
\begin{equation}
\rho_{\mathrm{g}}[\phi,\phi';t] \equiv \langle \phi| \hat
\rho_{\mathrm{g}}(t) |\phi'\rangle = {\cal N}\exp \! \left[\! - \!
\int \! \mathrm{d}^3x \mathrm{d}^3y\left( \phi(\vec x\,) A(\vec
x,\vec y;t)\phi(\vec y\,) +\phi'(\vec x\,) B(\vec x,\vec
y;t)\phi'(\vec y\,)-2\phi(\vec x\,) C(\vec x,\vec y;t)\phi'(\vec
y\,)\right) \right]\! , \label{density matrix: Gaussian}
\end{equation}
where $A,B,C$ and ${\cal N}$ are functions of the equal time field
correlators (for details see sections~\ref{Entropy in scalar field
theory}). Of course, the von Neumann equation~(\ref{density
op:eom}) does not hold any more for $\hat \rho_{\rm g}$, and thus
the entropy conservation law is violated by $\hat \rho_{\rm g}$.
Indeed, neglecting the information stored in observationally
inaccessible correlators will give rise to an increase of the
entropy\footnote{In passing we note that we shall not address
entropy generated due to the loss of unitarity caused by some
observer's inability to access the information stored in all of
the space-time. Prominent examples of this research field include
black hole and de Sitter space entropy and there is an extensive
literature on the subject. In \cite{Campo:2008ju} truncation of
the infinite hierarchy of Green's functions was proposed as an
operational definition of the entropy of cosmological
perturbations. Calzetta and Hu \cite{Calzetta:2003dk} coined the
expression ``correlation entropy'' and proved an H-theorem for the
correlation entropy in a quantum mechanical model.}.

The purpose of the present work is to calculate the entropy of a
system taking these considerations into account. We distinguish
two cases. In the simplest case we assume that our observer can
only measure Gaussian correlators and his apparatus is insensitive
to all non-Gaussian correlators. We will consider this situation
in sections \ref{Gaussian entropy from the Wigner function} and
\ref{Gaussian entropy from the replica trick}. In quantum field
theories, the problem seems to be that the evolution equations of
the three Gaussian correlators~(\ref{3 equal time correlators}) do
not close as soon as perturbative corrections are included. The
reason is that off-shell physics becomes important. A way out of
this impasse, and in our opinion {\it the} way out, is to solve
for the dynamics of the unequal time statistical correlator, i.e.
the statistical propagator, from which one then extracts the three
relevant correlators~(\ref{3 equal time correlators}).

In the second case, we assume that our observer is sensitive to
specific types of non-Gaussianities present in our theory, apart
from the leading order Gaussian correlators of course, either
because the observer has developed a sensitive measurement device
or the non-Gaussianities happen to be large enough. We will
consider this case in section \ref{Non-Gaussian entropy: Two
examples}. The sections above contain, for pedagogical reasons,
quantum mechanical derivations rather than quantum field
theoretical calculations. We will generalise our results to
quantum field theory in section \ref{Entropy in scalar field
theory}.

This paper aims only at developing the necessary machinery to
discuss entropy generation in an interacting quantum field theory
in an out-of-equilibrium setting, see e.g. \cite{Chou:1984es,
Jordan:1986ug, Calzetta:1986cq, Aarts:2001qa, Aarts:2001yn,
Aarts:2002dj, Berges:2002cz, Juchem:2003bi, Juchem:2004cs,
Berges:2004yj}. We discuss the application of these ideas
elsewhere \cite{Koksma:2009wa, Koksma:2010} in great detail, where
we consider a scalar field $\phi$ coupled to a second scalar field
via the interaction Lagrangian density ${\cal L}_{\rm int} =
-\frac{1}{2} h \chi^2(x) \phi(x)$. The results we develop here are
applicable to a much more wider class of field theoretical models,
including gauge and fermionic fields.

\section{Gaussian Entropy from the Wigner Function}
\label{Gaussian entropy from the Wigner function}

Entropy has of course been widely studied over the years
\cite{Wehrl:1978zz, Mackey:1989zz, Barnett:1991zz,
Serafini:2003ke, Cacciatori:2008qs, Adesso:2007jg}. In this
section we will derive an entropy formula based on a na\"ive
probabilistic interpretation of the Wigner function. This
characterises much of the 20th century efforts to connect the
Wigner function, Boltzmann's equations, Boltzmann's H-theorem and
macroscopic entropy \cite{Hillery:1983ms}. Let us begin by
considering the Hamiltonian of a time dependent harmonic
oscillator, which is the one-particle ``free'' Hamiltonian:
\begin{equation}\label{Hamiltonian:1particle}
H(p,x) = \frac{p^{2}}{2m} + \frac{1}{2} m \omega^{2} x^{2} +
H_{\mathrm{s}} \,, \qquad H_{\mathrm{s}}= x j \,.
\end{equation}
In general, the oscillator's mass and frequency and the source
current can all depend on time: $m=m(t)$, $\omega=\omega(t)$ and
$j=j(t)$, respectively. $H_{\mathrm{s}}$ is the source Hamiltonian
and $j=j(t)$ the corresponding current. This Hamiltonian is
relevant both for many condensed matter systems that can be
realised in laboratories and in an early Universe setting. For the
latter, the time dependence of the parameters is introduced for
example by the Universe's expansion or by phase transitions.
Alternatively, this Hamiltonian can represent for example the
simplest model of a laser, where $x=x(t)$ is the photon amplitude,
where, if $m \rightarrow 1$, $\omega(t) \rightarrow E(t)/\hbar$ is
the photon's frequency and where $j=j(t)$ is a charge current that
coherently pumps energy into the system, transforming the vacuum
state into a coherent state \cite{Glauber:1963tx,Cahill:1969iq}.

The time dependence of the parameters changes the system's energy,
but simultaneously does not change the Gaussian nature of a state
if such a state is imposed initially, i.e.: the evolution implied
by the Hamiltonian~(\ref{Hamiltonian:1particle}) transforms an
initial Gaussian density matrix into another Gaussian density
matrix. Moreover, as we will come to appreciate, one has to add a
coupling to the environment or add a self-interaction to generate
entropy. The Lagrangian $L = p \dot{x} - H $ follows as usual from
the Hamiltonian. Note that the source term can be removed from
this Lagrangian by a simple coordinate shift:
\begin{equation}\label{coordinateshift}
x(t) \rightarrow z(t) = x(t) - x_{0}(t) \,,
\end{equation}
upon which the Lagrangian reduces to a quadratic form:
\begin{equation}\label{Lagrangian:quadraticform}
L = \frac{1}{2} m\dot{z}^{2} - \frac{1}{2}m\omega^{2}
z^{2}+L_{0}(t) + \frac{\mathrm{d}}{\mathrm{d}t}\left[m z
\dot{x}_{0} \right]\,.
\end{equation}
Here, $L_{0}(t)$ is a $z$-independent function of time and the
last term is a boundary term which does not contribute to the
equation of motion if $x_{0}(t)$ is a function that obeys:
\begin{equation}\label{shift:x0}
\frac{\mathrm{d}}{\mathrm{d}t} \left[m\dot{x}_{0}
\right]+m\omega^{2} x_{0} + j = 0 \,.
\end{equation}
A general Gaussian density matrix centered at the new origin
remains a Gaussian density matrix centered at the origin under the
evolution of a quadratic Hamiltonian and moreover, which is what
the analysis above shows, linear source terms in the Hamiltonian
will not alter the Gaussian nature of the state. We can thus
consider Gaussian density matrices centered at the origin, whose
time evolution is governed by an Hamiltonian of the
form~(\ref{Hamiltonian:1particle}) with $j(t) \rightarrow 0$. When
written in position space representation, the single particle
density operator of a quantum mechanical Gaussian state centered
at the origin is of the form:
\begin{subequations}
\label{density operator: particle}
\begin{equation} \label{density operator: particle1}
\hat{\rho}_{\mathrm{g}}(t) = \int_{-\infty}^{\infty} \mathrm{d}x
\int_{-\infty}^{\infty} \mathrm{d}y |x\rangle
\rho_{\mathrm{g}}(x,y;t)\langle y |\,,
\end{equation}
where:
\begin{equation} \label{density operator: particle2}
\rho_{\mathrm{g}}(x,y;t) = {\cal N}(t) \exp\left[
-a(t)x^2-b(t)y^2+2c(t)xy \right] \,,
\end{equation}
\end{subequations}
and where $a=a(t)$, $b=b(t)$ and $c=c(t)$ are determined from the
von Neumann equation~(\ref{density op:eom}). The subscript g
denotes ``Gaussian''. Moreover, from $\hat
\rho_{\mathrm{g}}^\dagger = \hat\rho_{\mathrm{g}}$ it follows that
$b^* = a$ and $c^*=c$. When $c=0$ one recovers a pure state with
vanishing entropy. However when $c\neq 0$ the density matrix is
mixed and entangled and it cannot be written in the simple
diagonal form, $\rho_{\mathrm{g}}(x,y;t) = \Psi^*(y,t) \Psi(x,t)$,
where $\Psi(x,t) = \sqrt{{\cal N}}\exp(-ax^2)$. When $c>0$ the
density matrix tends to get diagonal in the $x-y$ direction,
however when $c<0$ the diagonalisation occurs in the $x+y$
direction in which case the entropy is not defined. The Heisenberg
uncertainty relation is in trouble in this case as we will see in
equation (\ref{uncertainly relation}). Hence it is natural to
assume that $c>0$. We keep the discussion phenomenological and do
not discuss the physical origin of $c\neq 0$. The normalisation
${\cal N}$ is obtained from requiring
$\mathrm{Tr}[\hat{\rho}_{\mathrm{g}}]=1$:
\begin{equation}\label{Norm of rho}
\mathrm{Tr}[\hat{\rho}_{\mathrm{g}}] = \int_{-\infty}^{\infty}
\mathrm{d} \tilde{x} \langle \tilde x| \hat \rho_{\mathrm{g}} |
\tilde x \rangle = \int_{-\infty}^{\infty} \mathrm{d}x
\rho_{\mathrm{g}}(x,x;t) = {\cal N}
\sqrt{\frac{\pi}{2(a_{\mathrm{R}}-c)}}= 1 \,,
\end{equation}
from which we conclude:
\begin{equation}\label{Norm of rho:2}
{\cal N} =  \sqrt{ \frac{2(a_{\mathrm{R}}-c)}{\pi}} \,,
\end{equation}
provided that $c < a_{\mathrm{R}}$ where $a_{\mathrm{R}}=\Re[a]$.
The equations that the functions $a(t)$, $b(t)$ and $c(t)$ of the
density matrix~(\ref{density operator: particle2}) obey can easily
be obtained from the von Neumann equation~(\ref{density op:eom}),
which in this Gaussian case in the amplitude basis reads:
\begin{equation} \label{density matrix:eom}
\imath\hbar\partial_t\rho_{\mathrm{g}}(x,y;t) = -
\frac{\hbar^2}{2m}\left(\partial^2_x-\partial^2_y\right)\rho_{\mathrm{g}}(x,y;t)
+ \frac12m\omega^2(x^2-y^2)\rho_{\mathrm{g}}(x,y;t) \,.
\end{equation}
If we insert equation (\ref{density operator: particle2}) in the
equation above we find:
\begin{subequations}
\label{density matrix:eom:2}
\begin{eqnarray}
\frac{\mathrm{d}a_{\mathrm{R}}}{\mathrm{d}t} &=&
\frac{4\hbar}{m}a_{\mathrm{I}} a_{\mathrm{R}} \label{density matrix:eom:2a}\\
\frac{\mathrm{d}c}{\mathrm{d}t} &=& \frac{4\hbar}{m}a_{\mathrm{I}}
c \label{density matrix:eom:2b}\\
\frac{\mathrm{d}}{\mathrm{d}t} \ln({\cal N}) &=&
\frac{2\hbar}{m}a_{\mathrm{I}} \label{density matrix:eom:2c}\\
\frac{\mathrm{d}a_{\mathrm{I}}}{\mathrm{d}t} &=&
\frac{2\hbar}{m}\left( a_{\mathrm{I}}^2 - a_{\mathrm{R}}^2 + c^2
\right) + \frac{m\omega^2}{2\hbar} \label{density matrix:eom:2d}
\,,
\end{eqnarray}
\end{subequations}
where we defined $a_{\mathrm{I}}=\Im[a]$. Note that ${\cal
N}\propto\sqrt{a_{\mathrm{R}}-c}$ is consistent with
equation~(\ref{density matrix:eom:2c}) as it
should\footnote{Indeed, by making use of
equations~(\ref{correlators:all}) one can show that
equations~(\ref{density matrix:eom:2}) are consistent with (and
equivalent to) the Hamilton equations for the
correlators~(\ref{Hamilton equations:correlators}).}.

The Wigner function is defined as a Wigner transform of the
density matrix $\rho_{\mathrm{g}}(x,y;t)$:
\begin{equation}\label{Wigner function:def}
{\cal W}(q,p;t) = \int_{-\infty}^{\infty} \mathrm{d} r {\rm
e}^{-\imath p r/\hbar}\rho_{\mathrm{g}}(q+r/2,q-r/2;t) \,,
\end{equation}
where we defined the average and relative coordinates, $q=(x+y)/2$
and $r = x-y$, respectively. Hence, a Wigner transform can be
thought of as an ordinary Fourier transform with respect to the
relative coordinate $r$ of the density matrix. The inverse Wigner
transform determines $\rho_{\mathrm{g}}(x,y;t)$ in terms of ${\cal
W}(q,p;t)$:
\begin{equation}\label{Wigner transform:inverse}
\rho_{\mathrm{g}}(x,y;t) = \int_{-\infty}^{\infty}
\frac{\mathrm{d} p}{2\pi\hbar}{\rm e}^{\imath p(x-y)/\hbar}{\cal
W}(q,p;t) \,,
\end{equation}
where $2\pi \hbar$ is the standard phase space measure. We obtain
the Gaussian Wigner function by performing the
integral~(\ref{Wigner function:def}):
\begin{equation}\label{Wigner function:Gauss}
{\cal W}(q,p;t) = {\cal M}(t) \exp\left[-\alpha_{\mathrm{w}}(t)
q^2 - \beta_{\mathrm{w}}(t) (p+q p_{\mathrm{c}}(t))^2 \right] \,,
\end{equation}
where:
\begin{subequations}
\label{Wigner function:Gauss2}
\begin{eqnarray}
\alpha_{\mathrm{w}} &=& 2(a_{\mathrm{R}}-c) \label{Wigner function:Gauss2a}\\
\beta_{\mathrm{w}} &=& \frac{1}{2\hbar^2 (a_{\mathrm{R}}+c)} \label{Wigner function:Gauss2b}\\
p_{\mathrm{c}}&=& 2\hbar a_{\mathrm{I}} \label{Wigner function:Gauss2c}\\
{\cal M} &=& \sqrt{\frac{4(a_{\mathrm{R}}-c)}{a_{\mathrm{R}}+c}}
\label{Wigner function:Gauss2d} \,,
\end{eqnarray}
\end{subequations}
and $p_{\mathrm{c}}$ denotes a ``classical'' momentum, which makes
the Wigner function non-diagonal. Note that the Wigner
function~(\ref{Wigner function:Gauss}) is normalised to unity, in
the sense that:
\begin{equation}\label{TrW=1}
\int_{-\infty}^{\infty} \frac{\mathrm{d}q\mathrm{d}p}{2\pi
\hbar}{\cal W}(q,p;t) = 1 \,.
\end{equation}
Recalling definition~(\ref{Wigner function:def}), equation
(\ref{density matrix:eom}) transforms to:
\begin{equation}
\Big(\partial_t + \frac{p}{m}\partial_q
                   - m\omega^2q\partial_p\Big) {\cal W}(q,p;t) = 0
\,, \label{Wigner function:eom}
\end{equation}
where we made use of the following relations:
\begin{subequations}
\label{intermediatestep1}
\begin{eqnarray}
\left(\partial_{x}^{2}-\partial_{y}^{2}\right)\rho_{\mathrm{g}}(x,y;t)
&=& 2
\partial_{q}\partial_{r} \rho_{\mathrm{g}}(q+r/2,q-r/2;t)
\label{intermediatestep1a}\\
r \mathrm{e}^{\imath p r/\hbar} &=& -\imath \hbar \partial_{p}
\mathrm{e}^{\imath p r /\hbar} \label{intermediatestep1b}\,.
\end{eqnarray}
\end{subequations}
Upon noting that $p/m = \dot x = \partial_p H$ and
$-m\omega^2x=\dot p =-\partial_x H$, we see that
equation~(\ref{Wigner function:eom}) is nothing but the Liouville
equation for the Boltzmann's distribution function $f(x,p;t)$,
\begin{equation}
\partial_t f(x,p;t) + \dot x \partial_x f(x,p;t) + \dot p \partial_p f(x,p;t) = 0
\label{Liouville}\,.
\end{equation}
A probabilistic interpretation of the Wigner function thus seems
plausible by the following identification:
\begin{equation}
         {\cal W}(q,p;t) \Leftrightarrow  f(x,p;t)
\,. \label{W=f}
\end{equation}
Of course, the equations for ${\cal W}(q,p;t)$ and $f(x,p;t)$ are
identical only in a free harmonic oscillator theory and, when
interactions are included, differences arise between the von
Neumann equation for the density operator $\hat
\rho_{\mathrm{g}}(t)$ (or the corresponding equation for ${\cal
W}(q,p;t)$) and the Boltzmann equation for $f(x,p;t)$. This indeed
makes the identification~(\ref{W=f}) less rigorous but it remains
a useful picture. The correspondence (\ref{W=f}) has been widely
used in the literature to develop a formalism relevant for example
for the physics of heavy ion collisions \cite{Mrowczynski:1992hq,
Aarts:2002dj} or for baryogenesis \cite{Prokopec:2003pj,
Prokopec:2004ic}.

In the distant 1969 Cahill and Glauber~\cite{Cahill:1969iq} have
proposed to approximate the density operator in the coherent state
basis $|\alpha\rangle$ by its diagonal form:
$\rho_{\mathrm{g}}\approx \rho_{\rm cs}(\alpha) = \langle
\alpha|\hat \rho_{\mathrm{g}}|\alpha\rangle$, which they interpret
as a quasi-probability distribution. Since coherent states harbour
many properties of classical particles, one can argue that this is
reasonable. Recall that coherent states are defined by $\hat
a|\alpha\rangle = \alpha|\alpha\rangle$, where $\hat a =
\sqrt{m\omega/2}[\hat x + \imath \hat p/(m\omega)]$ denotes the
annihilation operator for the
oscillator~(\ref{Hamiltonian:1particle}). The density matrix
$\rho_{\rm cs}(\alpha)$ represents an early example of a
quasi-probability distribution for a quantum system. Needless to
say, by this line of reasoning one can get only approximate
answers for the expectation value $\langle \hat A\rangle ={\rm
Tr}[\hat \rho_{\mathrm{g}}\hat A]$ of an operator $\hat A$ and
likewise for the entropy (which is just the expectation value of
$-\ln(\hat \rho_{\mathrm{g}})$). Another common way of
approximating expectation values of operators is to give a
probabilistic interpretation to the Wigner function, along the
same line of reasoning as argued above in equation~(\ref{W=f}) for
a free theory. In this approach~\cite{Cahill:1969iq,
Brandenberger:1992jh} the entropy is approximated by:
\begin{equation}
 S \approx S_{\cal W} \equiv - {\rm Tr}[{\cal W}(q,p;t) \ln({\cal W}(q,p;t))]
\,, \label{entropy:Wigner}
\end{equation}
where the trace ${\rm Tr}\rightarrow \int
\mathrm{d}p\mathrm{d}q/[2\pi \hbar]$ should be interpreted as an
integration over the phase space volume as in
equation~(\ref{TrW=1}). We now insert~(\ref{Wigner
function:Gauss}) into~(\ref{entropy:Wigner}) to find:
\begin{eqnarray}
S_{\cal W} &=& \frac12 \ln\left(\frac{a_{\mathrm{R}}
+c}{a_{\mathrm{R}}  -c}\right) + 1 -\ln(2)
 \nonumber\\
   &=& \ln\left(\frac{\Delta}{2}\right)  + 1
\,, \label{Wigner function:entropy:final}
\end{eqnarray}
where we defined:
\begin{equation}
\Delta^2  = \frac{a_{\mathrm{R}} +c}{a_{\mathrm{R}} -c} \,.
\label{Delta function:2}
\end{equation}
We evaluated the phase space Gaussian integrals by shifting the
momentum integration to $p'=p-p_c$ (the Jacobian of this shift
equals unity). At this point, it is useful to evaluate a few
quantum mechanical expectation values. We can extract the
following correlators from our Gaussian state:
\begin{subequations}
\label{correlators:all}
\begin{eqnarray}
\langle \hat x^2\rangle &=& {\rm Tr}[\hat\rho_{\mathrm{g}}\hat
x^2] = \int_{-\infty}^{\infty} \mathrm{d}\tilde{x} \langle
\tilde{x}| \hat\rho_{\mathrm{g}}\hat x^2 | \tilde{x}\rangle =
\frac{1}{4(a_{\mathrm{R}} -c)}
\label{correlators:alla}\\
\Big\langle \frac12\{\hat x,\hat p\}\Big\rangle &=&
-\hbar\frac{a_{\mathrm{I}} }{2(a_{\mathrm{R}} -c)} \label{correlators:allb}\\
\langle \hat p^2\rangle &=& \hbar^2\frac{|a|^2-c^2}{a_{\mathrm{R}}
-c} \,. \label{correlators:allc}
\end{eqnarray}
\end{subequations}
Here, we made use of $\langle x| \hat p |\psi\rangle = -
\imath\hbar \partial_{x} \langle x|\psi\rangle$. Moreover, one can
easily verify that $ \langle [\hat{x},\hat{p}]\rangle = \langle
\hat x\hat p-\hat p\hat x\rangle = \imath\hbar $ as it should.
This enables us to find the equivalent inverse relations:
\begin{subequations}
\label{aI:aR:c}
\begin{eqnarray}
\label{aI:aR:ca} a_{\mathrm{I}}  &=& -\frac{\left\langle
\frac12\{\hat x,\hat p\}\right\rangle}
                            {2\hbar\langle \hat x^2\rangle}
\label{aI:aR:cb}\\
a_{\mathrm{R}}  &=& \frac{\Delta^2+1}{8\langle \hat x^2\rangle}
\label{aI:aR:cc}\\
c &=& \frac{\Delta^2-1}{8\langle \hat x^2\rangle} \,,
\label{aI:aR:cd}
\end{eqnarray}
\end{subequations}
and to express $\Delta(t)$ in equation (\ref{Delta function:2}) in
terms of the correlators~(\ref{correlators:all}):
\begin{equation}
\Delta^2(t) = \frac{4}{\hbar^2}
      \left[
         \langle \hat x^2\rangle \langle \hat p^2\rangle
          -  \Big\langle \frac12\{\hat x,\hat p\}\Big\rangle^2
      \right]
\,. \label{Delta function:3}
\end{equation}
For the moment this is just a definition, but as we will come to
discuss, the physical meaning of $\hbar\Delta(t)$ is the phase
space area in units of $\hbar$ occupied by a (Gaussian) state
centered at the origin. For a pure state $\Delta(t) = 1$, while
for a mixed state $\Delta(t)>1$. Hence, we can define the
uncertainty relation for a general Gaussian state:
\begin{equation}
 \frac{(\hbar\Delta(t))^2}{4} = \langle \hat x^2\rangle \langle \hat p^2\rangle
          -  \Big\langle \frac12\{\hat x,\hat p\}\Big\rangle^2
          \geq \frac{\hbar^2}{4}
\,. \label{uncertainly relation}
\end{equation}
If $c$ is negative and $|c|< a_{\mathrm{R}}$, we can in principle
have $\Delta <1$, which can be seen from equation (\ref{Delta
function:2}). It would be interesting to investigate whether any
physical experiment could be proposed where a violation of
Heisenberg's uncertainty relation occurs, even if it were for only
a very brief moment in time (such that it would still hold on
average).

Furthermore, it is natural to define the statistical particle
number density on our phase space, in the quantum field
theoretical sense, as:
\begin{equation}
 n(t) \equiv \frac{\Delta(t)-1}{2}
\,, \label{particle number}
\end{equation}
in terms of which the Wigner entropy~(\ref{Wigner
function:entropy:final}) can be rewritten as:
\begin{equation}
S_{\cal W}(t) = \ln\left(n(t)+\frac 12\right)  + 1 \,.
\label{Wigner function:entropy:2}
\end{equation}
These results in essence agree with the entropy per mode of scalar
cosmological perturbations obtained in
\cite{Brandenberger:1992jh}. Shortly we shall see that the Wigner
entropy~(\ref{Wigner function:entropy:2}) is a good approximation
for the von Neumann entropy if $n\gg 1$, or equivalently when
$\hbar \Delta \gg 1$.

Let us try to put these results briefly in historical context. The
fundamental question that is left unanswered
in~\cite{Brandenberger:1992jh} is what the dynamical mechanism for
the growth of $\Delta$ is. Instead, the authors argue that an
observer that measures cosmological perturbations can be
approximated by a coherent state (which is a good classical
basis), thus explaining the growth in $\Delta$. Two mechanisms
were considered in \cite{Prokopec:1992ia}: a projection on the
particle number basis and on the coherent state basis. In both
cases, information of the phase of the squeezed Gaussian state is
lost, which thus generates entropy. In the latter approach
formula~(\ref{Delta function:3}) is reproduced, while in the
former, the entropy acquires the form of the statistical entropy
of Bose particles $S=(n+1)\ln(n+1)-n\ln(n)$ with $n\rightarrow
\sinh^2(r)$ the particle number associated with the squeezed
state, and $r$ the squeezing parameter of the state. While these
early articles, and many subsequent ones~\cite{Polarski:1995jg,
Kiefer:1999sj}, attempted to explain the origin of entropy of
cosmological perturbations by a (non-unitary) process of coarse
graining (associated with particular properties of the observer's
measuring apparatus), a genuine dynamical mechanism of entropy
generation was lacking.

This has changed recently \cite{Campo:2004sz, Campo:2005sy,
Prokopec:2006fc, Kiefer:2006je, Kiefer:2007zza, Campo:2008ju,
Campo:2008ij, Koksma:2009wa}, now dynamical mechanisms for entropy
generation have been proposed. Rather than discussing various
mechanisms in detail, let us just point out that the notion of how
entropy should be defined and generated has evolved: recently, it
has been advocated that the relevant entropy to consider is the
Gaussian entropy, although the approaches differ in the actual
origin of the entropy increase. For example, Prokopec and
Rigopoulos suggested in~\cite{Prokopec:2006fc} that tracing over
the unobserved isocurvature perturbations leads to entropy
generation in the adiabatic mode. Similarly, tracing over the
unobserved gravitational waves should lead to entropy generation
in the adiabatic mode~\cite{Koksma:2007zz}. Kiefer, Lohmar,
Polarski and Starobinsky~\cite{Kiefer:2006je,Kiefer:2007zza} as
well as Campo and Parentani~\cite{Campo:2008ju,Campo:2008ij},
among others, suggested that, if super-Hubble modes take the role
of the system, the origin of the entropy of cosmological
perturbations should be associated with the stochastic noise
generated by the sub-Hubble ultraviolet modes representing the
environment. Both Giraud and Serrau \cite{Giraud:2009tn} and we
\cite{Koksma:Talk,Koksma:Talk2,Koksma:2009wa} advocate that
neglecting observationally inaccessible non-Gaussianities is a
prominent mechanism to generate entropy.
\begin{figure}[t!]
    \begin{minipage}[t]{.43\textwidth}
        \begin{center}
\includegraphics[width=\textwidth]{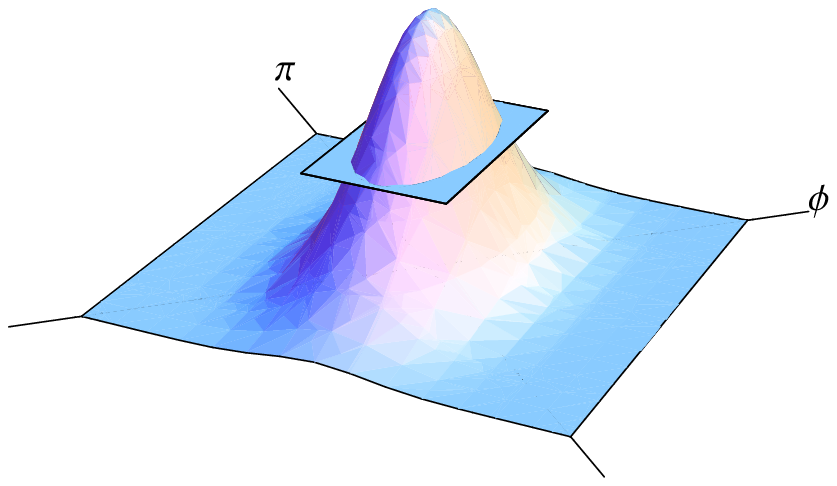}
   {\em \caption{General shape of the Wigner function,
   representing the 2-dimensional Gaussian phase space distribution of a
   certain state. The $1\sigma$ or $2\sigma$ cross-section is an
   ellipse. For the current quantum mechanical discussion, we of course let
   $\pi\rightarrow p$ and $\phi \rightarrow q$.
    \label{fig:GaussianPhaseSpace3}}}
        \end{center}
    \end{minipage}
\vskip .1 cm
    \begin{minipage}[t]{.43\textwidth}
        \begin{center}
\includegraphics[width=\textwidth]{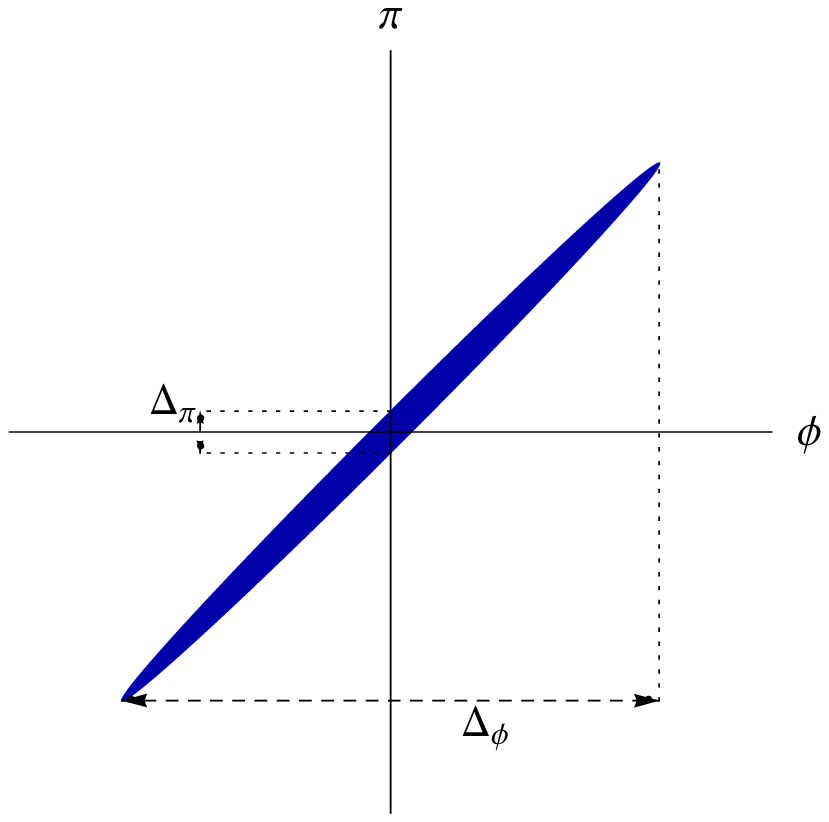}
   {\em \caption{Squeezed state before decoherence. For example, the area of the ellipse of a vacuum state would be unity $\Delta=1$. For the current quantum mechanical discussion, we of course let
   $\pi\rightarrow p$ and $\phi \rightarrow q$, also in figure \ref{fig:GaussianPhaseSpace2}. \label{fig:GaussianPhaseSpace1}}}
        \end{center}
    \end{minipage}
\hfill
\begin{minipage}[t]{.43\textwidth}
        \begin{center}
\includegraphics[width=\textwidth]{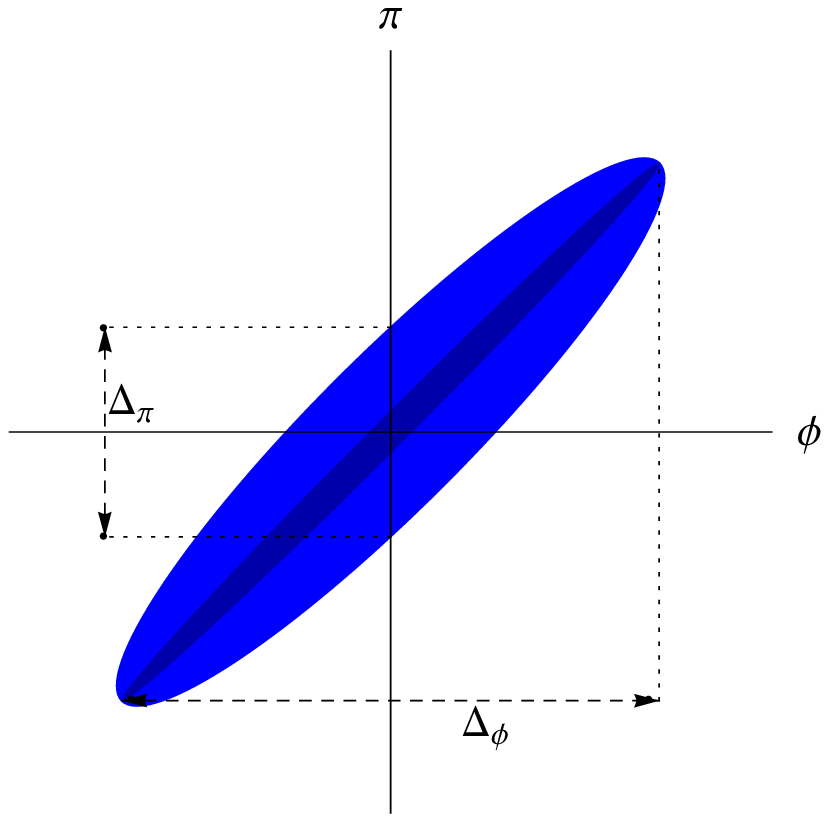}
   {\em \caption{Squeezed state after decoherence. Clearly, the
   accessible phase space has increased and one can say that
   knowledge about the system has been lost. The entropy has
   increased compared to figure \ref{fig:GaussianPhaseSpace1}.
    \label{fig:GaussianPhaseSpace2}}}
        \end{center}
    \end{minipage}
\end{figure}

So far, we have discussed how the Gaussian entropy in Wigner space
is defined and qualitatively discussed various mechanisms to
generate entropy dynamically. Let us now, to finish this section,
discuss how one can sensibly think of a decohered state in Wigner
space and discuss the physical meaning of $\Delta(t)$. Despite the
fact that the Wigner entropy (\ref{Wigner function:entropy:final})
only approximates the von Neumann entropy, it offers a useful,
intuitive way of depicting decohered states irrespective of the
precise underlying mechanism by which such a state has decohered
in the first place. Let us, for this reason, discuss this
phenomenologically. The Wigner function (\ref{Wigner
function:Gauss}) is a 2-dimensional Gaussian function whose width
in the $p$-direction need a priori not necessarily be the same as
in the $q$-direction. Hence, in general it can be squeezed in some
direction. This is illustrated in
figure~\ref{fig:GaussianPhaseSpace3}. The $1\sigma$ or $2\sigma$
cross-section of the phase space distribution has the shape of an
ellipse, which we also visualised in figure
\ref{fig:GaussianPhaseSpace3}. If the cross-section is a circle
centered at the origin, the state is either pure or mixed and if,
moreover, its area is unity $\Delta(t)=1$ it is a pure vacuum
state. If the cross-section is again a circle but displaced from
the origin, we are dealing with a generalised coherent-squeezed
state (whose shape is not described by equation (\ref{Wigner
function:entropy:final}) for obvious reasons). If the
cross-section is an ellipse, it is a squeezed state as mentioned
before. The ellipse is parametrised by:
\begin{equation}\label{Wignerslice}
\alpha_{\mathrm{w}}(t)q^{2}+\beta_{\mathrm{w}}(t)\left(p+q \,
p_{\mathrm{c}}(t)\right)^{2} = \vartheta\,,
\end{equation}
where we have made use of equation (\ref{Wigner function:Gauss})
and where $\vartheta$ is some constant that determines the height
at which we slice the Wigner function. The area of this ellipse is
now given by\footnote{Note that the area of ellipse defined by the
equation $Ax^{2}+Bxy+Cy^{2}=1$ is
$\mathcal{A}=2\pi/\sqrt{4AC-B^2}$.}:
\begin{equation}\label{Wignerarea}
\mathcal{A}(t) = \frac{\vartheta
\pi}{\sqrt{\alpha_{\mathrm{w}}(t)\beta_{\mathrm{w}}(t)}} =
\vartheta \pi \hbar \Delta(t) \,,
\end{equation}
where we have used equations (\ref{Wigner function:Gauss2}) and
(\ref{Delta function:2}). Clearly, the area of the ellipse in
Wigner space is determined straightforwardly from the phase space
area $\Delta(t)$. Note that if $\vartheta = 1/\pi$ we have
$\mathcal{A} = \Delta$.

Figures \ref{fig:GaussianPhaseSpace1} and
\ref{fig:GaussianPhaseSpace2} visualise the decoherence process of
a squeezed state. Initially, in figure
\ref{fig:GaussianPhaseSpace1}, the state is squeezed and has unity
phase space area $\Delta(t_{0})=1$, such that its entropy
vanishes. Now we switch on our favourite decoherence mechanism:
the state interacts with some environment and environmental
degrees of freedom are traced over or non-Gaussian correlators
generated by the interaction with the environment are neglected.
Consequently, the area in phase space increases $\Delta(t)>1$,
such that $S(t)>0$. Hence, the ellipse in Wigner space grows which
we depict in figure \ref{fig:GaussianPhaseSpace2}. An important
feature of any decoherence process reveals itself: for a highly
squeezed state, given the knowledge of $q$ after some measurement,
the value $p$ can take in a subsequent measurement is constraint
in a very narrow interval. Interestingly, after the state has
decohered, and given the same measurement of $q$, the range of
values $p$ can take has increased. Indeed one can say that
knowledge about $p$ has been lost compared to the squeezed state
before decoherence, hence entropy has been generated.
Alternatively, we can say that given some measurement of $q$, the
value $p$ can take is, after decoherence, drawn from a classical
stochastic distribution, which indeed coincides with the familiar
idea that decohered quantum systems should behave as uncorrelated
stochastic systems. Indeed, $n$ roughly counts the number of
patches in phase space that behave independently and are
uncorrelated.

With this intuitive notion of decohered state in mind, one can
understand that, for highly squeezed states, the position space
basis (which generalises to the field amplitude basis in the
quantum field theoretical case) is the pointer basis. Let us start
out with a highly squeezed state, i.e.: take for example a pure
state during inflation that rapidly squeezes during the Universe's
expansion. The Hamiltonian of the state is then dominated by the
potential term, and the kinetic term only contributes little. In
thermal equilibrium, the system will minimise its free energy
$F=H-TS$, where $T$ is the temperature of the heat bath. Now, if
we switch on some decoherence mechanism due to interaction with
the environment, the entropy will increase mainly due to momentum
increase, i.e.: $\langle \hat{p}^2 \rangle$ will increase. Indeed,
increasing $\langle \hat{p}^2 \rangle$ will hardly affect the
Hamiltonian, whereas it will significantly affect the entropy.
Hence, the variable $x$ is robust during the process of
decoherence, in the sense that $\langle \hat{x}^{2} \rangle$ will
hardly change such that it qualifies as a proper pointer basis.
Note however, that $x$ is only a pointer basis in the statistical
sense, such that there is a well defined probability distribution
function from which a measurement can be drawn.

\section{Gaussian entropy from the replica trick}
\label{Gaussian entropy from the replica trick}

The entropy of a general Gaussian state has been derived by Sohma,
Holevo and
Hirota~\cite{SohmaHolevoHirota:1999,SohmaHolevoHirota:2000} by
making use of Glauber's P representation for the density operator.
Here we present an alternative derivation for the von Neumann
entropy~(\ref{entropy:vN}) of a general Gaussian state by making
use of the replica trick (see e.g.:
\cite{Callan:1994py,Calabrese:2009qy,Calabrese:2004eu}). Of
course, there are other analogous methods one can use\footnote{We
thank Theo Ruijgrok for his useful comments.}. Notice first that:
\begin{equation}
  \ln(\hat\rho_{\mathrm{g}}) = \lim_{\epsilon\rightarrow 0}
              \frac{\hat\rho_{\mathrm{g}}^\epsilon-1}{\epsilon}
              \,.
\label{replica trick}
\end{equation}
Here, $\epsilon$ is a positive integer which is analytically
extended to zero. Hence, in order to calculate the
entropy~(\ref{entropy:vN}) we need to evaluate ${\rm
Tr}\left[\hat\rho_{\mathrm{g}}^{1+\epsilon}\right]$. Using
equation (\ref{density operator: particle}), we thus have:
\begin{eqnarray}
{\rm Tr}\left[\hat\rho_{\mathrm{g}}^{1+\epsilon}\right]
  &=& {\cal N}^{1+\epsilon}
      \int_{-\infty}^{\infty}  \mathrm{d}x_0 \cdots \int_{-\infty}^{\infty} \mathrm{d}x_{\epsilon}\exp \left[-2a_{\mathrm{R}} \left(x_0^2+\cdots +x_\epsilon^2 \right)
                    +2c \left(x_0x_1 + x_1x_2 + \cdots + x_\epsilon x_0 \right) \right]
\nonumber\\
 &=& {\cal N}^{1+\epsilon}\int_{-\infty}^{\infty} \mathrm{d}x_0 \cdots \int_{-\infty}^{\infty} \mathrm{d}x_{\epsilon}\exp
 \left[-2\beta \left\{ (x_0 - \alpha x_1)^2+(x_1 - \alpha x_2)^2+ \cdots +
                (x_\epsilon - \alpha x_0)^2 \right\} \right]
 \nonumber\\
 &=& |1-\alpha^{1+\epsilon}|^{-1}
       \left({\cal N}\int_{-\infty}^{\infty} \mathrm{d}y \exp\left[ -2\beta y^2 \right] \right)^{1+\epsilon}
 \nonumber\\
 &=&  |1-\alpha^{1+\epsilon}|^{-1}
             \left(\frac{a_{\mathrm{R}}-c}{\beta}\right)^{\frac{1+\epsilon}{2}}\,.
\label{replica trick:2}
\end{eqnarray}
Here, $J=|1-\alpha^{1+\epsilon}|^{-1}$ is the Jacobian of the
transformation from $x_i$ to $y_i=x_i-\alpha x_{i+1}$, where
$i=\{0,1,\cdots ,\epsilon-1\}$, and $y_\epsilon=x_\epsilon-\alpha
x_{0}$. A comparison of the first and the second line
in~(\ref{replica trick:2}) tells us that:
\begin{subequations}
\label{replica trick:3}
\begin{eqnarray}
\alpha_\pm &=&
\frac{a_{\mathrm{R}}}{c}\pm\sqrt{\left(\frac{a_{\mathrm{R}}}{c}\right)^2-1}
\label{replica trick:3a}\\
\beta_\pm &=& \frac{c}{2\alpha_\pm} = \frac{c\alpha_\mp}{2} \,.
\label{replica trick:3b}
\end{eqnarray}
\end{subequations}
In the limit $c\rightarrow 0$ one must recover a pure diagonal
density operator and, consequently, the shift $\alpha$ should
vanish, singling out:
\begin{subequations}
\label{replica trick:4}
\begin{eqnarray}
\alpha &=& \alpha_- = \frac{a_{\mathrm{R}}}{c} -
\sqrt{\left(\frac{a_{\mathrm{R}}}{c}\right)^2-1}
\label{replica trick:4a}\\
\beta &=& \beta_- \,. \label{replica trick:4b}
\end{eqnarray}
\end{subequations}
in equation~(\ref{replica trick:3}) as the physical choice. Based
on $(a_{\mathrm{R}}-c)/\beta=(1-\alpha)^2$, we can rewrite
equation~(\ref{replica trick:2}) in a simpler form:
\begin{equation}
{\rm Tr}\left[\hat\rho_{\mathrm{g}}^{1+\epsilon}\right]
   =  \frac{(1-\alpha)^{1+\epsilon}}{|1-\alpha^{1+\epsilon}|}
\,. \label{replica trick:5}
\end{equation}
Recall that normalisability of the density operator requires
$a_{\mathrm{R}}>c$, implying that $0\leq\alpha<1$, such that the
absolute value in (\ref{replica trick:5}) can be dropped. This
means that ${\rm
Tr}\left[\hat\rho_{\mathrm{g}}^{1+\epsilon}\right]$ is analytic in
$\alpha$ (for positive integers $\epsilon$):
\begin{equation}
{\rm Tr}\left[\hat\rho_{\mathrm{g}}^{1+\epsilon}\right]
   =  \frac{(1-\alpha)^{1+\epsilon}}{1-\alpha^{1+\epsilon}}
\,, \label{replica trick:6}
\end{equation}
such that one can analytically extend to complex $\epsilon$.
According to the replica method, the entropy is then obtained by
taking the limit $\epsilon\rightarrow 0$. To perform this
procedure we expand~(\ref{replica trick:6}) to linear order in
$\epsilon$ and we find the following expression for the von
Neumann entropy~(\ref{entropy:vN}):
\begin{equation}
S_{\rm vN}(t) = -\ln(1-\alpha)-\frac{\alpha}{1-\alpha}\ln(\alpha)
\,. \label{entropy:rho log rho:final}
\end{equation}
Using the expression for $\alpha$ in equation~(\ref{replica
trick:4a}) and the definition of the phase space area in equations
(\ref{Delta function:2}) and~(\ref{Delta function:3}), we find:
\begin{equation}
\alpha(t) = \frac{\Delta(t)-1}{\Delta(t)+1} \,,
\label{alphatoDelta}
\end{equation}
such that we can express the entropy~(\ref{entropy:rho log
rho:final}) solely in terms of $\Delta$:
\begin{equation}
S_{\rm vN}(t) =
\frac{\Delta+1}{2}\ln\left(\frac{\Delta+1}{2}\right) -
\frac{\Delta-1}{2}\ln\left(\frac{\Delta-1}{2}\right)\,.
\label{entropy:final}
\end{equation}
Since $0\leq\alpha(t)<1$, we have $1 \leq \Delta(t)  < \infty$.
The von Neumann entropy~(\ref{entropy:final}) vanishes when
$\Delta=1$ (or $c=0$), defining a pure state, while a strictly
positive entropy $S_{\rm vN}>0$ implies a mixed state $\Delta >
1$. Clearly, the Gaussian von Neumann entropy is completely
determined by the three Gaussian correlators characterising the
state (\ref{Delta function:3}). Relations~(\ref{entropy:final})
and~(\ref{entropy:rho log rho:final}) suggest the following
definition of the particle number $n(t)$:
\begin{equation}
n(t) \equiv \frac{\alpha(t)}{1-\alpha(t)} = \frac{\Delta(t) -1}{2}
\,, \label{n}
\end{equation}
where $0\leq n<\infty$, in terms of which the
entropy~(\ref{entropy:rho log rho:final}), (\ref{entropy:final})
becomes:
\begin{equation}
S_{\rm vN}(t)=(1+n)\ln(1+n)-n\ln(n) \,, \label{entropy:n}
\end{equation}
which is the well known result from statistical physics for the
entropy of $n$ Bose particles per (quantum)
state~\cite{Kubo:1965}. Note that (\ref{entropy:n}) is a convex
function of $n$, such that $S(n_1)+S(n_2)
> S(n_1+n_2)$, which is another desirable property for the entropy. The particle number $n$ defined
in~(\ref{n}) should be interpreted as the number of independent
(uncorrelated) regions in the phase space of a single (one
particle) quantum state.

We will show next that $\Delta$ is a special function, as it is
conserved by the evolution equations resulting from the
Hamiltonian~(\ref{Hamiltonian:1particle}):
\begin{subequations}
\label{Hamilton equations}
\begin{eqnarray}
\dot{\hat x} &=& \partial_p \hat H = \frac{\hat p}{m}
\label{Hamilton equationsa} \\
\dot{\hat p} &=& -\partial_x \hat H = - m\omega^2\hat x \,,
\label{Hamilton equationsb}
\end{eqnarray}
\end{subequations}
where we set $j\rightarrow 0$. These equations imply for the
correlators:
\begin{subequations}
\label{Hamilton equations:correlators}
\begin{eqnarray}
\frac{\mathrm{d}}{\mathrm{d}t}\langle\hat x^2\rangle
    &=& \frac{2}{m}\Big\langle\frac12\{\hat x,\hat p\}\Big\rangle
\label{Hamilton equations:correlatorsa}\\
\frac{\mathrm{d}}{\mathrm{d}t}\Big\langle\frac12\{\hat x,\hat
p\}\Big\rangle
     &=& - m\omega^2\langle\hat x^2\rangle
         + \frac{1}{m}\langle\hat p^2\rangle
\label{Hamilton equations:correlatorsb}\\
\frac{\mathrm{d}}{\mathrm{d}t}\langle[\hat x,\hat p]\rangle &=& 0
\label{Hamilton equations:correlatorsc}\\
\frac{\mathrm{d}}{\mathrm{d}t}\langle\hat p^2\rangle
     &=& - 2m\omega^2\Big\langle\frac12\{\hat x,\hat p\}\Big\rangle
\label{Hamilton equations:correlatorsd}\,.
\end{eqnarray}
\end{subequations}
Recall that these equations are, in the light of
equation~(\ref{aI:aR:c}), equivalent to~(\ref{density
matrix:eom:2}). The third equation is automatically satisfied by
the commutation relations, $[\hat x,\hat p] = \imath \hbar$. One
can combine the other three equations in~(\ref{Hamilton
equations:correlators}) to show that:
\begin{equation}
  \frac{\mathrm{d}}{\mathrm{d}t}  \frac{\hbar^2\Delta^2}{4}
 = \frac{\mathrm{d}}{\mathrm{d}t}\left[\langle\hat x^2\rangle \langle\hat p^2\rangle
              - \Big\langle\frac12\{\hat x,\hat p\}\Big\rangle^2\right]
 = 0
\label{Delta:conservation}
\end{equation}
This implies that $\Delta$ (and in fact any function of $\Delta$)
is conserved by the Hamiltonian evolution~(\ref{Hamilton
equations}--\ref{Hamilton equations:correlators}).

Of course, when interactions are included, the conservation law
(\ref{Delta:conservation}) becomes approximate. In quantum field
theory interactions are typically cubic or quartic in the fields
and will thus induce non-Gaussianities in the density matrix. In
particular, for a quantum mechanical $\lambda\phi^4$ model, this
has been investigated by Calzetta and Hu \cite{Calzetta:2002ub,
Calzetta:2003dk}. They derive an H-theorem for $\Delta$ in the
case of a quantum mechanical $O(N)$ model.

\begin{figure}[t!]
    \begin{minipage}[t]{.43\textwidth}
        \begin{center}
\includegraphics[width=\textwidth]{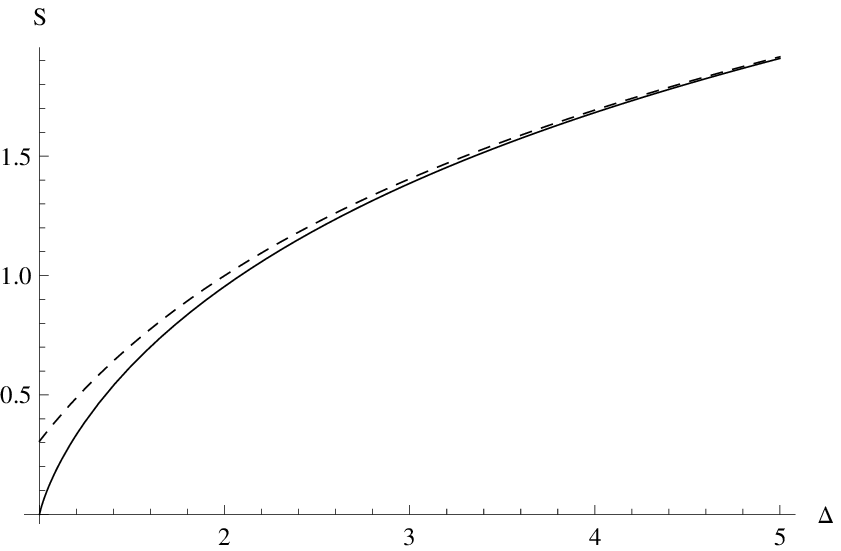}
         \end{center}
    \end{minipage}
\hfill
    \begin{minipage}[t]{.43\textwidth}
        \begin{center}
\includegraphics[width=\textwidth]{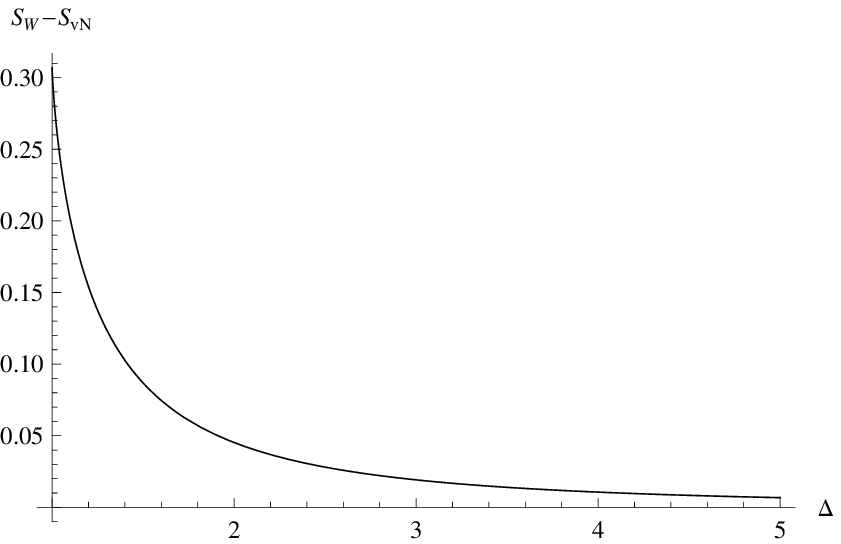}
        \end{center}
    \end{minipage}
\begin{minipage}{.86\textwidth}
\begin{center}
{\em \caption{Left: The von Neumann entropy (solid line) and its
Wigner entropy approximation (dashed line) for a Gaussian state as
a function of the phase space area $\Delta$. Right: The difference
between the Wigner and the von Neumann entropy. The Wigner entropy
approaches the exact von Neumann entropy rapidly as $\Delta$
grows. \label{fig:entropycomp}}}
\end{center}
\end{minipage}
\end{figure}
Finally, it is interesting to compare the von Neumann
entropy~(\ref{entropy:final}) and the Wigner entropy~(\ref{Wigner
function:entropy:final}), as shown in
figure~\ref{fig:entropycomp}. This reveals that the Wigner entropy
represents a large phase space (semiclassical) limit of the von
Neumann entropy, which should not come as a surprise. Indeed, the
Wigner entropy can be used to accurately represent the entropy in
systems which develop large correlators (occupation numbers), such
as cosmological perturbations~\cite{Brandenberger:1992jh}. The
Wigner entropy fails however to take account of quantum
correlations present in the density matrix and Wigner function,
which are properly taken into account in the exact expression for
the entropy of a general Gaussian state~(\ref{entropy:final}).

\section{Non-Gaussian entropy: Two examples}
\label{Non-Gaussian entropy: Two examples}

An important question is how to generalise the von Neumann entropy
of a Gaussian state to include non-Gaussian corrections.
Non-Gaussianities can be important either because they are made
large on purpose by specially preparing the state, or simply
because they are accessible in measurements due to the fact that
the observer's measurement device is very accurate.

While we are far from building a general theory of how
non-Gaussian correlations affect the notion of phase space volume,
statistical particle number and thus entropy, we shall present two
examples in this section: we consider a system in a state which
can be represented by an admixture of a Gaussian density matrix
and some small non-Gaussian contributions. Secondly, we consider
small quartic corrections to the density matrix, contributing to
the kurtosis of the ground state.

Qualitatively, we expect that the correction to the Gaussian
entropy due to non-Gaussianities present in a theory is small,
except for systems whose entropy is (almost) zero. This can most
easily be appreciated in the Wigner approach to entropy. Consider
for example a system with a large entropy, for example a squeezed
state whose phase space area in Wigner space has increased
significantly due to developing a $\Delta \gg 1$.
Non-Gaussianities can deform the area in phase space of the state
(by squeezing, pushing or stretching the Wigner function) roughly
by a factor of order unity. Consequently, the effect on the
entropy will be relatively small. This can be understood by
realising that a lot of information is contained in unequal time
correlators. Of course, such an argument ceases to be true for a
pure state, whose quantum properties are very pronounced.

\subsection{Admixture of the Ground State and First Excited State} \label{Admixture of the Ground State and First Excited State}

Let us consider a density matrix of the form:
\begin{equation}
 \rho_2(x,y;t) = {\cal N}_2(1+ \zeta_0 x + \zeta_0^*y+\zeta_1 x^2 + \zeta_1^*y^2+2\zeta_2 xy)
               \exp\left(-ax^2-a^*y^2+2cxy\right)
\,. \label{rho2}
\end{equation}
Such a state typically appears when one uses a laser to excite an
harmonic oscillator in its ground state. The system can then be
described by an admixture of the ground state and a small
contribution to the first exited state\footnote{Recall that the
wave function of the pure one particle state of the simple
harmonic oscillator~(\ref{Hamiltonian:1particle}) (with $j=0$) is:
\begin{equation}
\psi_1(x) = \langle x|\psi_1 \rangle     =
\bigg[\frac{4}{\pi}\bigg(\frac{m\omega}{\hbar}\bigg)^3\bigg]^{1/4}
           \, x\, \exp\Big(-\frac{m\omega x^2}{2\hbar}\Big)
\,, \nonumber
\end{equation}
with energy $E_1 = \hbar \omega\left(1+\frac12\right)$. The
problem at hand reduces to this state upon identifying
$a_{\mathrm{R}} \leftrightarrow m\omega/(2\hbar)$,
$a_{\mathrm{I}}\rightarrow 0$, $c\rightarrow 0$,
$\zeta_0\rightarrow 0$ and $\zeta_1\rightarrow 0$.}. Note that the
process of pumping energy in a simple harmonic oscillator in its
ground state could lead to creating a state that is not pure ($c
\neq 0$). The measuring apparatus is assumed to be sensitive to
the admixture of the two states. We shall assume that
non-Gaussianity is small, in the sense that all $\zeta_i$
parameters ($i=0,1,2$) are small, such that we content ourselves
with performing the analysis up to linear order in $\zeta_i$.

Notice first that to this order the terms in parentheses in
equation~(\ref{rho2}) can be approximated by an exponential:
\begin{equation}
1+ \zeta_0 x + \zeta_0^*y+\zeta_1 x^2 + \zeta_1^*y^2+2\zeta_2 xy
\simeq \exp\left(\zeta_0 x + \zeta_0^*y+\zeta_1 x^2 +
\zeta_1^*y^2+2\zeta_2 xy\right) \,, \label{exponentialassumption}
\end{equation}
implying that the first two terms $\zeta_0 x+\zeta_0^*y$ induce an
entropy conserving shift in $x$ and $y$ (as in coherent states).
Of course, this ceases to be true at quadratic and higher orders
in $\zeta_0$. On the other hand, the $\zeta_1$ and $\zeta_2$ terms
do change the entropy even at linear order. At higher order,
$\zeta_1$ and $\zeta_2$ induce kurtosis whose effect on the
entropy we will discuss shortly.

Before we turn our attention to calculating the entropy, let us
first show that a density matrix (\ref{rho2}) with $\zeta_0
\rightarrow 0$ indeed generates non-trivial higher order
correlators. Normalising the trace to unity yields:
\begin{equation}
{\cal N}_2 = \sqrt{\frac{2}{\pi}}
\frac{(a_{R}-c)^{\frac{3}{2}}}{a_{\mathrm{R}}+c+\frac{1}{2}(\zeta_{1R}+\zeta_2)}
 \,. \label{normdensity2}
\end{equation}
An interesting higher order correlator to consider is for example
the connected part of the four-point correlator:
\begin{equation}
\langle \hat x^{4} \rangle_{\mathrm{con}} = \langle \hat x^{4}
\rangle - 3(\langle \hat x^{2} \rangle)^{2} \,.
\label{4pntcorrelator}
\end{equation}
If the connected part of the four-point correlator is non-zero,
the state is said to have kurtosis. Using the Gaussian density
matrix in equation (\ref{density operator: particle}) one can
easily show that the connected part of the four-point correlator
vanishes as it should: for free theories higher order correlators
either vanish or can be expressed in terms of Gaussian
correlators. Using however the non-Gaussian density matrix in
equation (\ref{rho2}) with $\zeta_0 \rightarrow 0$ and the
normalisation constant (\ref{normdensity2}) we find:
\begin{equation}
\langle \hat x^{4} \rangle_{\mathrm{con}} = -
\frac{3(\zeta_{1R}+\zeta_2)^{2}}{16(a_{\mathrm{R}}-c)^{2}(a_{\mathrm{R}}-c+\frac{1}{2}(\zeta_{1R}+\zeta_2))^{2}}
\simeq -
\frac{3(\zeta_{1R}+\zeta_2)^{2}}{16(a_{\mathrm{R}}-c)^{4}}
+\mathcal{O}\left((\zeta_{1R})^{3},(\zeta_2)^{3}\right) \,.
\label{4pntcorrelator2}
\end{equation}
Clearly, genuine non-Gaussian correlators are generated and this
state has kurtosis at quadratic order.

Let us now calculate the entropy by making use of equation
(\ref{exponentialassumption}). As said, it is precisely
$\zeta_{1}$ and $\zeta_{2}$ that affect the entropy. To see that
notice further that their effect can be captured by a shift in $a$
and $c$ as follows:
\begin{subequations}
\label{a+c:rescaling}
\begin{eqnarray}
a &\rightarrow & \bar a = a - \zeta_1 \label{a+c:rescalinga}\\
c &\rightarrow & \bar c = c + \zeta_2 \label{a+c:rescalingb}\,,
\end{eqnarray}
\end{subequations}
such that the corresponding von Neumann entropy simply reads ({\it
cf.} equation (\ref{entropy:n})):
\begin{equation}
S_2 = (1+\bar n)\ln(1+\bar n)-\bar n\ln(\bar n) \label{s2}
\,,
\end{equation}
where $\bar n = n + \delta n$ and where as usual $n=(\Delta
-1)/2$. Now $\delta n\propto \zeta_{1R},\zeta_2$ is the correction
to the Gaussian entropy we are about to calculate. Comparing with
equations~(\ref{replica trick:3}) and~(\ref{n}) we have:
\begin{subequations}
\label{nandalphabar}
\begin{eqnarray}
\bar \alpha &=& \frac{\bar a_{\mathrm{R}}}{\bar c} -
\sqrt{\frac{\bar
a_{\mathrm{R}}^2}{\bar c^2}-1} \label{nandalphabara}\\
\bar n &=& \frac{\bar \alpha}{1-\bar\alpha}
\label{nandalphabarb}\,,
\end{eqnarray}
\end{subequations}
from which we immediately find, to linear order in $\zeta_{1R}$
and $\zeta_{2}$:
\begin{subequations}
\label{deltananddeltaalpha}
\begin{eqnarray}
\delta\alpha &=& \frac{\alpha(1-\alpha)}{1+\alpha}
\left(\frac{\zeta_{1R}}{a_{\mathrm{R}}-c}+
\frac{1+\alpha^2}{2\alpha}\frac{\zeta_2}{a_{\mathrm{R}}-c}\right)
\label{deltananddeltaalphaa} \\
\delta n &=& \frac{\delta\alpha}{(1-\alpha)^2}
          = \frac{n(1+n)}{1+2n}
\left(\frac{\zeta_{1R}}{a_{\mathrm{R}}-c}+
\frac{1+2n+2n^2}{2n(1+n)}\frac{\zeta_2}{a_{\mathrm{R}}-c}\right)
\label{deltananddeltaalphab}\,,
\end{eqnarray}
\end{subequations}
where we used $a_{\mathrm{R}}/c = (1+\alpha^2)/(2\alpha)$.
Finally, $\delta S_{2} = \{\ln[(1+n)/n]\}\delta n$ implies that:
\begin{subequations}
\label{s2:final}
\begin{equation}
S_2(t)=S_{\mathrm{g}}(t) + \delta S_{2}(t) \label{s2:finala} \,,
\end{equation}
where $\delta S_{2}$ is an entropy shift given by:
\begin{equation}
\delta S_2 (t)=  \frac{n(1+n)}{1+2n}
\left(\frac{\zeta_{1R}}{a_{\mathrm{R}}-c}+
\frac{1+2n+2n^2}{2n(1+n)}\frac{\zeta_2 }{a_{\mathrm{R}}-c}\right)
\ln\left(\frac{1+n}{n}\right) \label{s2:finalb}\,,
\end{equation}
\end{subequations}
where $S_{\rm g}=(1+n)\ln(1+n)-n\ln(n)$ is the von Neumann entropy
of a Gaussian state and $n$ is the statistical particle number
associated with the Gaussian part of the state as before.

Several comments are in order. Firstly, the
result~(\ref{s2:final}) implies that to linear order
$\Im[\zeta_1]$ does not change the entropy. In fact,
$\Im[\zeta_1]$ induces (de)squeezing of the state, which can be
appreciated from equation (\ref{Wigner function:Gauss2c}).
Secondly, $\delta S_2$ is positive (negative) whenever
$\zeta_{1R}$ and $\zeta_2$ are positive (negative), irrespective
of $n>0$. Thirdly, we have rescaled $\zeta_{1R}$ and $\zeta_2$ by
$(a_{\mathrm{R}}-c)$ to get a dimensionless quantity. This
rescaling is natural, since $a_{\mathrm{R}}-c$ measures the width
of the state. Finally, for large $n$ the formula~(\ref{s2:final})
gives a meaningful answer, since:
\begin{equation}
\lim_{n\rightarrow \infty} \delta S_2 =
\frac{1}{2}\frac{\zeta_{1R}+\zeta_2}{a_{\mathrm{R}}-c}\,,
\label{Entropylimit1}
\end{equation}
is finite. On the other hand, when $n\rightarrow 0$, we encounter
a logarithmic divergence:
\begin{equation}
\lim_{n\rightarrow 0} \delta S_2 =
\left(n\frac{\zeta_{1R}}{a_{\mathrm{R}}-c}+
\frac{1}{2}\frac{\zeta_2}{a_{\mathrm{R}}-c}\right)
\ln\Big(\frac{1}{n}\Big)\,, \label{Entropylimit2}
\end{equation}
indicating a mild (logarithmic) breakdown of the linear expansion.
Notice however that also in that limit the formula~(\ref{s2})
remains applicable.

\subsection{Kurtosis}
\label{Kurtosis}

In the former example non-Gaussianities are generated by adding a
quadratic polynomial to the prefactor. One could also introduce
non-Gaussianities by adding higher order, i.e.: cubic or quartic,
powers in the exponential. Cubic corrections generate skewness and
since they contribute to the entropy at quadratic and higher
orders in the skewness parameter, we shall now focus on the
quartic corrections to the density matrix. Quartic corrections to
a density matrix affect the kurtosis of a state and we parametrise
this as follows:
\begin{equation}
 \rho_4(x,y;t) = {\cal N}_4\exp \left(-ax^2-a^*y^2+2cxy \right)\exp \left(\eta_0x^4+\eta_0^*y^4+\eta_1x^3y+\eta_1^*xy^3+\eta_2x^2y^2 \right)
   \equiv \frac{{\cal N}_4}{{\cal N}_{\rm g}} \rho_{\rm g} (x,y;t) \rho_{\rm
   ng}(x,y;t),
 \label{rho4}
\end{equation}
where $\rho_{\rm g}$ is the Gaussian density matrix (\ref{density
operator: particle}) as before, with the normalisation ${\cal
N}_{\rm g}$ given in equation (\ref{Norm of rho:2}), and where
finally $\rho_{\rm
ng}=\exp(\eta_0x^4+\eta_0^*y^4+\eta_1x^3y+\eta_1^*xy^3+\eta_2x^2y^2)$.
As in the previous example, we shall consider only linear
corrections in $\eta_i$ ($i=0,1,2$) to the entropy. To this order
the non-Gaussian part of~(\ref{rho4}) can also be written as:
\begin{equation}
\delta \rho_{\rm ng} \equiv \rho_{\rm ng} - 1 =
\eta_0x^4+\eta_0^*y^4+\eta_1x^3y+\eta_1^*xy^3+\eta_2x^2y^2 +{\cal
O}(\eta_i^2) \,. \label{rhong:lin}
\end{equation}
We can find the normalisation constant after some simple algebra
by making use of~(\ref{rhong:lin}):
\begin{equation}
 {\cal N}_{4} = {{\cal N}_{\rm g}}
  \bigg[1+\frac34 \frac{\eta_0+\eta_0^*+\eta_1+\eta_1^*+\eta_2}{[2(a_{\mathrm{R}}-c)]^2}
  \bigg]^{-1}
\,. \label{calN4}
\end{equation}

\begin{figure}[t!]
    \begin{minipage}[t]{.43\textwidth}
        \begin{center}
\includegraphics[width=\textwidth]{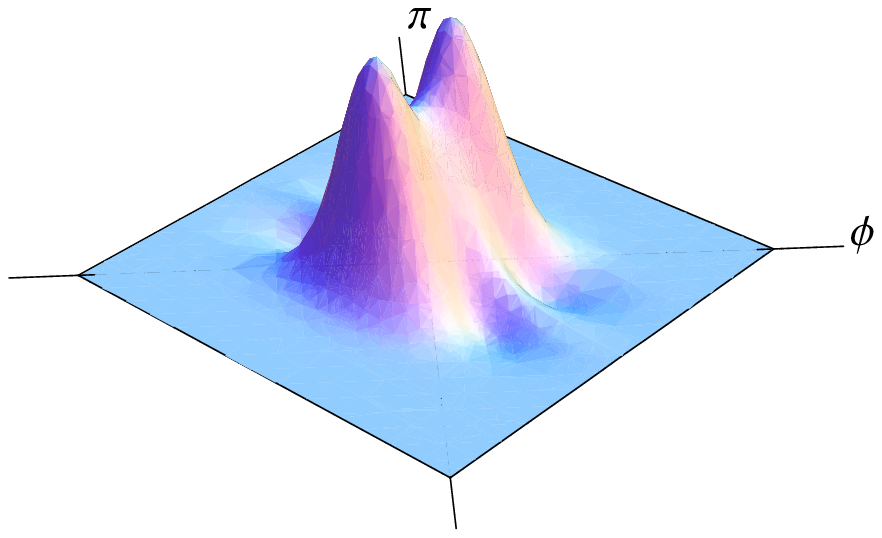}
         \end{center}
    \end{minipage}
\hfill
    \begin{minipage}[t]{.43\textwidth}
        \begin{center}
\includegraphics[width=\textwidth]{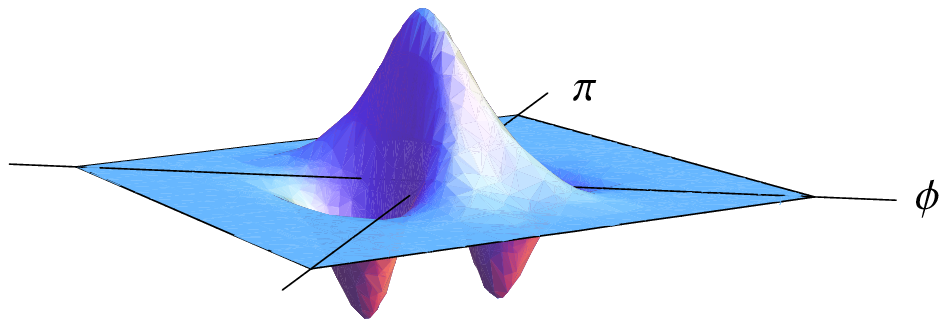}
        \end{center}
    \end{minipage}
\begin{minipage}{.86\textwidth}
\begin{center}
{\em \caption{Wigner transform of the density matrix described
   by equations (\ref{rho4}) and (\ref{rhong:lin}), with $\eta_0
   \neq 0$ only. We used $\eta_0>0$ (left) and $\eta_0<0$ (right). The peak structure of our state in phase space, that was so simple for Gaussian
   density matrices, is much more complicated. Some regions in
   phase space have a negative value of the
   Wigner function, indicating a breakdown of the use of a Wigner
   function as a convenient measure of probability.
    \label{fig:NGWigner}}}
\end{center}
\end{minipage}
\end{figure}
To gain some intuitive understanding of what our density matrix
looks like, let us consider figure \ref{fig:NGWigner}. Here, we
show the Wigner transform of equation (\ref{rho4}) using
(\ref{rhong:lin}), with $\eta_1 = 0 = \eta_2$. Clearly, the
quartic corrections change the peak structure of our state in
phase space. Moreover, some regions in phase space now have a
negative Wigner function. This nicely illustrates the limitations
of using the Wigner function as a probability density on our phase
space as soon as non-Gaussianities (due to interactions) are
included, i.e.: equation (\ref{W=f}) holds only approximately in
this case.

In order to calculate the entropy, we need ${\rm Tr}[\hat
\rho_4^{1+\epsilon}]$ as before ({\it cf.} equation~(\ref{replica
trick:2})). To linear order in $\delta \rho_{\rm ng}$ we have:
\begin{equation}
{\rm Tr}[\hat \rho_4^{1+\epsilon}] = \left(\frac{{\cal N}_4}{{\cal
N}_{\rm g}}\right)^{1+\epsilon}
     \Big( {\rm Tr}[\hat \rho_{\rm g}^{1+\epsilon}]
  + (1+\epsilon){\rm Tr}[\hat \rho_{\rm g}^{1+\epsilon}\delta\hat\rho_{\rm ng}]
    \Big)
\,. \label{rho4:lin}
\end{equation}
The first term in the parentheses is just the Gaussian
result~(\ref{replica trick:2}) while the latter term can be
evaluated by making use of the formulae in Appendix \ref{Appendix
A Useful integrals}. Expanding to linear order in $\epsilon$, the
equation above yields:
\begin{eqnarray}
{\rm Tr}[\hat \rho_4^{1+\epsilon}]
 &=& 1 + \epsilon\Big[\ln(1-\alpha)+\frac{\alpha}{1-\alpha}\ln(\alpha)\Big]
 +\epsilon
 \bigg[
      \frac34\frac{\eta_0+\eta_0^*}{[2(a_{\mathrm{R}}-c)]^2}
                 \Big(\frac{4\alpha}{1-\alpha^2}\ln(\alpha)-1\Big)
 \nonumber\\
 &&\hskip 1cm
     +\,\frac34\frac{\eta_1+\eta_1^*}{[2(a_{\mathrm{R}}-c)]^2}
                 \Big(\frac{1+\alpha}{1-\alpha}\ln(\alpha)-1\Big)
     +\frac34\frac{\eta_2}{[2(a_{\mathrm{R}}-c)]^2}
                 \Big(\frac43\frac{1+\alpha+\alpha^2}{1-\alpha^2}\ln(\alpha)-1\Big)
 \bigg]
\,.\qquad \label{rho4:lin:b}
\end{eqnarray}
As expected, we see that the entropy naturally splits into a
Gaussian and a non-Gaussian contribution:
\begin{subequations}
\label{deltaS4}
\begin{equation}
  S_4 (t)=  S_{\rm g}(t) +\delta S_4(t)
\label{deltaS4a}\,,
\end{equation}
where $S_{\rm g}$ is the Gaussian entropy as before, and where:
\begin{eqnarray}
 \delta S_4(t) &=&
      \frac{\eta_0 +\eta_0^* }{[2(a_{\mathrm{R}}\!-\!c)]^2}
        \left(\frac{3n(1+n)}{1+2n}\ln\left(\frac{1+n}{n}\right)+\frac34\right)
     +\frac34\frac{\eta_1+\eta_1^*}{[2(a_{\mathrm{R}}\!-\!c)]^2}
                 \left((1+2n)\ln\left(\frac{1+n}{n}\right)+1\right)
\nonumber\\
    && +\,\frac{\eta_2}{[2(a_{\mathrm{R}}\!-\!c)]^2}
                 \left( \frac{1+3n+3n^2}{1+2n}\ln\left(\frac{1+n}{n}\right)+\frac34\right)
 \,.
\label{deltaS4b}
\end{eqnarray}
\end{subequations}
This is the main result of this section and intuitive in the
following sense: just as in the first non-Gaussian example above,
we see that positive kurtosis parameters
$\{\eta_{0R},\eta_{1R},\eta_2\}>0$, tend to make the effective
state's width $(a_{\mathrm{R}}-c)^{-1/2}$ larger, which increases
the area in phase space the state occupies, which in turn
increases the entropy. If however
$\{\eta_{0R},\eta_{1R},\eta_2\}<0$, the phase space area shrinks,
which in turn decreases the entropy. The statements above hold for
any statistical particle number $0\ll n<\infty$, with the
exception of $n\rightarrow 0$, where, just as in the case studied
above, a weak logarithmic divergence occurs when $\eta_{1R}\neq 0$
or $\eta_2\neq 0$. Notice finally that $\Im[\eta_0]$ and
$\Im[\eta_1]$ again do not participate in entropy generation, but
rather contribute to the squeezing of the state.

Kurtosis and skewness in quantum mechanics, studied in this
section, occur also in interacting quantum field theories which we
discuss next.

\section{Entropy in Scalar Field Theory}
\label{Entropy in scalar field theory}

The quantum mechanical expressions for the Gaussian and
non-Gaussian entropies that we have developed for pedagogical
reasons in sections \ref{Gaussian entropy from the Wigner
function}, \ref{Gaussian entropy from the replica trick}
and~\ref{Non-Gaussian entropy: Two examples} for a single particle
density matrix can be generalised to field theory. We firstly need
to consider correlators in quantum field theory however.

\subsection{Equal Time Correlators in Scalar Field Theory}
\label{Equal Time Correlators in Scalar Field Theory}

Let us now proceed analogous to equation~(\ref{density operator:
particle}) and write the density matrix operator for our system in
the field amplitude basis in Schr\"odinger's picture (see e.g.
\cite{Koksma:2007uq}):
\begin{equation}
\label{density operator: field} \hat \rho_{\mathrm{g}} = \int
{\cal D}\phi \int {\cal D}\phi^\prime
    |\phi\rangle  \rho_{\mathrm{g}}[\phi,\phi^\prime;t]\langle \phi^\prime| \,,
\end{equation}
where:
\begin{equation}
\label{density operator: field2}
\rho_{\mathrm{g}}[\phi,\phi^\prime;t] = {\cal N}
\exp\left(-\phi^T\!\cdot A\cdot\phi - {\phi^\prime}^T\!\cdot
B\!\cdot\!\phi^\prime + 2\phi^T\!\cdot C\!\cdot\!\phi^\prime
\right) \,,
\end{equation}
where ${\cal D}\phi = \prod_{\vec x\in V}\mathrm{d}\phi(\vec x\,)$
and $|\phi\rangle = \prod_{\vec x\in V}|\phi(\vec x\,)\rangle $. A
few words on the notation first. We shall consider $\phi=\phi(\vec
x)$ as a vector whose components are labelled by $\vec x$.
Moreover, $A(\vec x,\vec y,t)$ can be viewed as a matrix, such
that $A\cdot \phi$ is a vector again, where a $\cdot$ denotes
matrix multiplication which, for the case at hand, is nothing but
an integral over $D-1$ dimensional space. Hence, quantities like
$\phi^{\mathrm{T}}\cdot A \cdot \phi$ are scalars and involve two
integrals over space. Note that at this point we do not assume
that $A$ is homogeneous, i.e.: $A(\vec x,\vec y,t) \neq A(\vec x -
\vec y,t)$, but rather keep any possible off-diagonal terms for
generality.

Our density matrix (\ref{density operator: field2}) is hermitian,
such that we have $A^*(\vec x,\vec y,t)=B(\vec x,\vec y,t)$ and
$C^*(\vec x,\vec y,t)=C(\vec x,\vec y;t)$, where we also used
$A(\vec x,\vec y,t)=A(\vec y,\vec x,t)$ and $C(\vec x,\vec
y,t)=C(\vec y,\vec x;t)$. The normalisation ${\cal N}$ of
$\rho[\phi,\phi';t]$ can be determined from the standard
requirement ${\rm Tr}[\hat\rho]=1$:
\begin{equation}
{\cal N} = \left({\rm
det}\left[\frac{2(A_{\mathrm{R}}-C)}{\pi}\,\right]\right)^{1/2}
\,, \label{normrhofunc}
\end{equation}
where we note that the hermitian part $A_{\mathrm{h}}$ of $A$ is
real and symmetric such that $A_{\mathrm{h}}=A_{\mathrm{R}}$.

Just as in the quantum mechanical case, we need to calculate the
three non-trivial Gaussian correlators which completely
characterise the properties of our Gaussian state. In order to
calculate these correlators, we need to have an expression for the
statistical propagator as in equation (\ref{3 equal time
correlators}) which, given some initial density matrix
$\hat{\rho}(t_0)$, is in the Heisenberg picture defined by:
\begin{equation}
F(\vec x,t;\vec y,t')={\rm Tr}\left[\hat\rho(t_0)\hat\phi(\vec
x,t)\hat\phi(\vec y,t')\right] \label{statisticalpropagator} \,.
\end{equation}
Let us begin by calculating:
\begin{equation}
\langle \hat\phi(\vec x) \hat\phi(\vec y) \rangle = F(\vec x,\vec
y;t)={\rm Tr}\left[\hat\rho(t)\hat\phi(\vec x\,)\hat\phi(\vec
y\,)\right] \label{F:1} \,,
\end{equation}
where we have made use of the Heisenberg evolution equation for
operators. It is convenient to add a source current $j(\vec x,t)$
to the density matrix~(\ref{density operator: field}), such that
$\rho$ becomes:
\begin{equation}
\rho_{\mathrm{g}}^{j}[\phi,\phi^\prime;t] = {\cal N}
\exp\left(-\phi^T\!\cdot A\cdot\phi - {\phi^\prime}^T\!\cdot
A^{\dag}\!\cdot\!\phi^\prime + 2\phi^T\!\cdot
C\!\cdot\!\phi^\prime + j^T\!\cdot \phi + {j'}^T\!\cdot
\phi^\prime \right) \,, \label{density operator: field:j}
\end{equation}
in terms of which equation (\ref{F:1}) can be rewritten as:
\begin{subequations}
\label{correlatorsQFT}
\begin{eqnarray}
F(\vec x,\vec y;t) &=&\frac{\delta}{\delta j(\vec x\,)}
\frac{\delta}{\delta j(\vec y\,)} \int {\cal D}\phi
\rho^j_{\mathrm{g}}[\phi,\phi;t]\Big|_{j=j^\prime=0}
\nonumber\\
&=&\frac{\delta}{\delta j(\vec x\,)}\frac{\delta}{\delta j(\vec
y\,)}\int {\cal D}\tilde\phi \, {\cal N} \left.
\exp\left(-2\tilde\phi^T\!\cdot(A_{\mathrm{R}}-C)\cdot\tilde
\phi\right)\exp\left(\frac18(j+j^\prime)^T\!\cdot(A_{\mathrm{R}}-C)^{-1}
\cdot(j+j^\prime)\right)  \right|_{j=j^\prime=0}
\nonumber\\
&=& \frac14(A_{\mathrm{R}}-C)^{-1}(\vec x,\vec y;t)\,.
\label{correlatorsQFTa}
\end{eqnarray}
We also need the other correlators:
\begin{eqnarray}
\frac{1}{2} \langle \{ \hat{\phi}(\vec{x}), \hat{\pi}(\vec{y}) \}
\rangle &=& \partial_{t'} F(\vec x,t;\vec
y,t^\prime)|_{t=t^\prime} = \frac12{\rm Tr}[\hat
\rho_{\mathrm{g}}(t) \{\hat\phi(\vec x\,),\hat\pi(\vec y\,)\}]
=-\frac{\hbar}{2}(A_{\mathrm{R}}-C)^{-1}\cdot A_{\mathrm{I}}(\vec
x,\vec y;t)
\label{correlatorsQFTb} \\
\frac{1}{2} \langle \{ \hat{\pi}(\vec{x}), \hat{\pi}(\vec{y}) \}
\rangle &=&  \partial_t\partial_{t^\prime}F(\vec x,t;\vec
y,t^\prime)|_{t=t^\prime} = \frac12{\rm Tr}[\hat\rho_{\mathrm{g}}
(t) \{\hat\pi(\vec x\,),\hat\pi(\vec y\,)\}]
\label{correlatorsQFTc}\\
&=& \hbar^2 \! \left[ \frac{1}{2} A^{\dag} \! \cdot\!
(A_{\mathrm{R}} \! -C)^{-1} \! \cdot \!A \!+\! A \cdot\!
(A_{\mathrm{R}}\!-C)^{-1} \! \cdot A^{\dag}\! -C \!\cdot\!
(A_{\mathrm{R}}\!-C)^{-1}\! \cdot C \!\right](\vec x,\vec y;t)
\nonumber\,.
\end{eqnarray}
\end{subequations}
We have moreover made use of $\langle\phi'|\hat{\pi}|\phi\rangle =
-\imath \hbar \frac{\delta}{\delta\phi'} \delta[\phi'-\phi]$. As a
check one can verify that:
\begin{equation}
{\rm Tr}(\hat \rho_{\mathrm{g}}(t)[\hat\phi(\vec x\,),\hat\pi(\vec
y\,)]) = \imath\hbar\delta^{\scriptscriptstyle{D}-1}(\vec x-\vec
y\,) \,. \label{commutator}
\end{equation}
Combining the equations above we find:
\begin{eqnarray}
\Delta^2(\vec x,\vec y;t) &=& \frac{4}{\hbar^2} \left[ \left
\langle\hat\phi\hat\phi\right\rangle \cdot \left
\langle\hat\pi\hat\pi \right \rangle- \frac14 \left\langle
\{\hat\phi,\hat\pi\}\right\rangle \cdot\left\langle
\{\hat\phi,\hat\pi\}\right\rangle \right](\vec x,\vec y;t)
\nonumber\\
         &=& (A_{\mathrm{R}}-C)^{-1}\cdot(A_{\mathrm{R}}+C)(\vec x,\vec y;t)
\,. \label{Delta function:phi}
\end{eqnarray}
This is the desired field theoretic generalisation of the Gaussian
invariant $\Delta^2$ in equation (\ref{Delta function:3}). Notice
that the result above applies for general non-diagonal Gaussian
density matrices. This is an important quantity because, just as
in the quantum mechanical case, $\Delta$ will be the conserved
quantity under any quadratic Hamiltonian evolution. Since the von
Neumann entropy is also conserved in this case, it is natural to
expect that $S_{\mathrm{vN}}=S_{\mathrm{vN}}[\Delta]$.

Of course, if we are interested in problems in which the
hamiltonian density is only time dependent, one can make use of
spatial translation invariance of the correlators, such that the
equal time statistical correlator is homogeneous: $F(\vec x, \vec
y;t) \rightarrow \tilde F(\vec x-\vec y,t)$. In this case it is
beneficial to Fourier transform according to:
\begin{equation}
\label{Fouriertransform} \tilde F (\vec k,t) = \int
\mathrm{d}^{\scriptscriptstyle{D}-1}(\vec x - \vec y)
 \tilde  F(\vec x-\vec y,t) \mathrm{e}^{-\imath\vec
k\cdot(\vec x - \vec y)}\,,
\end{equation}
such that equation (\ref{Delta function:phi}) becomes local in
momentum space and reduces to the result known in the literature:
\begin{eqnarray}
\tilde{\Delta}^2(\vec k;t) &=& \frac{4}{\hbar^2} \left.
\left[\tilde F(\vec k,t,t)
\partial_t\partial_{t^\prime}\tilde F(\vec k,t,t^\prime) -
\left(\partial_{t'} \tilde F(\vec k,t,t^\prime) \right)^2 \right]
\right|_{t=t^\prime}
\nonumber\\
&=&\frac{(\tilde A_{\mathrm{R}}+ \tilde C)(\vec k,t)}{( \tilde
A_{\mathrm{R}}- \tilde C)(\vec k,t)} \,. \label{Delta
function:phi:momentum}
\end{eqnarray}
This representation is particularly useful in problems with
spatial translational symmetry, such as
cosmology~\cite{Brandenberger:1992jh,Campo:2008ij}.

\subsection{Entropy of a Gaussian state in Scalar Field Theory}
\label{Entropy of a Gaussian state in scalar field theory}

Let us now discuss the von Neumann entropy~(\ref{entropy:vN}) of
the Gaussian density matrix~(\ref{density operator: field}) by
using the replica method~(\ref{replica trick}). One can proceed
analogous to section~\ref{Gaussian entropy from the replica
trick}. Some subtleties arise however as we deal with a system
with infinite degrees of freedom. For this reason, we nevertheless
include an outline of the proof in appendix~\ref{Appendix B
Entropy for Gaussian Field Theory}. The entropy for a quantum
system that can be described by a Gaussian density matrix is given
by:
\begin{equation}
S_{\rm vN} = {\rm
Tr}\left[\frac{\Delta+\mathbb{I}}{2}\cdot\ln\left(\frac{\Delta+\mathbb{I}}{2}\right)
- \frac{\Delta-\mathbb{I}}{2}\cdot
\ln\left(\frac{\Delta-\mathbb{I}}{2}\right) \right] \,.
\label{entropy:vN:Delta}
\end{equation}
We denote the identity matrix by
$\mathbb{I}=\delta^{\scriptscriptstyle{D}-1}(\vec x -\vec y)$. As
in the quantum mechanical case, we can define the generalised
statistical particle number density correlator $n=n(\vec x,\vec
y,t)$ as:
\begin{equation}
n = \frac{\Delta -\mathbb{I}}{2} \label{n:matrix}\,,
\end{equation}
in terms of which the entropy~(\ref{entropy:vN:Delta}) reads:
\begin{equation}
S_{\rm vN} =  {\rm Tr}\left[(n+\mathbb{I})\cdot\ln
\left(n+\mathbb{I} \right) - n\cdot\ln \left(n \right) \right] \,.
\label{entropy:vN:n}
\end{equation}
In the limit when $\|n \| \gg 1$, in the sense that for the
diagonalised matrix its diagonal entries of interest are large, we
can expand the logarithms in~(\ref{entropy:vN:n}) to get for the
entropy, $S_{\rm vN} \approx {\rm Tr}[\ln(n)+\mathbb{I} +{\cal
O}(n^{-1})]$, which nearly coincides with the Wigner entropy
$S_{\cal W}$ in field theory, {\it cf.} equation~(\ref{Wigner
function:entropy:2}). Equations~(\ref{entropy:vN:n})
and~(\ref{entropy:vN:Delta}) represent the von Neumann entropy of
a general Gaussian state in scalar field theory and are the main
result of this section.

In the homogeneous limit, we can again Fourier transform and
equation (\ref{entropy:vN:Delta}) reduces to:
\begin{equation}
S_{\rm vN} = V \int \frac{\mathrm{d}^{3}\mathbf{k}}{(2\pi)^3}
\left[\frac{\Delta (\mathbf{k},t)+1}{2}
\ln\left(\frac{\Delta(\mathbf{k},t)+1}{2}\right) -
\frac{\Delta(\mathbf{k},t)-1}{2}
\ln\left(\frac{\Delta(\mathbf{k},t)-1}{2}\right) \right] \,,
\label{entropy:vN:Delta_FourierTransform}
\end{equation}
where the volume factor arises because of the trace. The limit
$\Delta(\mathbf{k},t) \gg 1$ basically agrees with
\cite{Brandenberger:1992jh}.

\subsection{Entropy of a Non-Gaussian state in Scalar Field Theory}
\label{Entropy of simple non-Gaussian states in scalar field
theory}

Analogous to the one particle non-Gaussian entropy discussed in
section~\ref{Non-Gaussian entropy: Two examples}, let us
generalise that result to the field theoretical case. The
non-Gaussian density matrix $\hat \rho_2$ in
equation~(\ref{density operator: field}) generalises to:
\begin{subequations}
\label{density operator: field:nonG}
\begin{equation}
\hat \rho_2 = \int {\cal D}\phi \int {\cal D}\phi^\prime
|\phi\rangle  \rho_2[\phi,\phi^\prime;t]\langle \phi^\prime| \,,
\label{density operator: field:nonGa}
\end{equation}
where:
\begin{eqnarray}
\rho_2[\phi,\phi^\prime;t] &=& {\cal N}_2\left[1+\zeta_0\cdot\phi
+ \zeta_0^\dagger\cdot\phi^\prime +\phi^T\cdot\zeta_1\cdot\phi +
{\phi^\prime}^T\cdot\zeta_1^\dagger\cdot\phi^\prime
+\phi^T\cdot\zeta_2\cdot{\phi^\prime} +
{\phi^\prime}^T\cdot\zeta_2^\dagger\cdot\phi
\right]\rho_0[\phi,\phi^\prime;t] \label{density operator:
field:nonGb}
\\
\rho_0[\phi,\phi^\prime;t] &=& \exp\left[-\phi^T\!\cdot A\cdot\phi
                    - {\phi^\prime}^T\!\cdot A^{\dag}\!\cdot\!\phi^\prime
                    + \phi^T\!\cdot C\!\cdot\!\phi^\prime
                    + {\phi^\prime}^T\!\cdot C^T\cdot\!\phi
                \right] \,,
\label{density operator: field:nonGc}
\end{eqnarray}
\end{subequations}
We could of course perform a similar shift in the exponent as we
have done before in equation (\ref{a+c:rescaling}) in which case
the resulting equation for the entropy would formally be exact,
i.e: we do not assume yet that the non-Gaussian contributions are
small. If we want to expand around the Gaussian result however, we
can only perform the integral if we assume that all correlators,
including the non-Gaussian ones, are homogeneous, i.e.: they are
only a function of the difference of their coordinates $A(\vec
x,\vec y,t)=A(\vec x-\vec y,t)$. The Fourier transform of equation
(\ref{density operator: field:nonG}) is:
\begin{equation}
\rho_2 [\phi,\phi^\prime;t] = \prod_{\mathbf{k}} {\cal
N}_{2,\mathbf{k}} \exp\left[- \bar
A(\mathbf{k},t)|\phi(\mathbf{k})|^{2}- \bar
A^{\ast}(\mathbf{k},t)|\phi'(\mathbf{k})|^{2} + 2 \bar
C(\mathbf{k},t)\phi^{\ast}(\mathbf{k}) \phi'(\mathbf{k}) \right] +
{\cal O}(\zeta_i^2) \,, \label{density operator: field:nonG2}
\end{equation}
where we have set $\zeta_0=0$ as before, as it will not induce an
entropy shift. Also, the product over the momenta is only over
half of the Fourier space. Moreover, we have absorbed the small
non-Gaussian contributions in the functions $\bar A$ and $\bar C$
as in the quantum mechanical case:
\begin{subequations}
\label{a+c:rescalingQFT}
\begin{eqnarray}
A(\mathbf{k},t) &\rightarrow & \bar A(\mathbf{k},t) = A(\mathbf{k},t) - \zeta_1 (\mathbf{k},t) \label{a+c:rescalingQFTa}\\
C(\mathbf{k},t) &\rightarrow & \bar C(\mathbf{k},t) =
C(\mathbf{k},t) + \zeta_2 (\mathbf{k},t)
\label{a+c:rescalingQFTb}\,,
\end{eqnarray}
\end{subequations}
Now we can read off the result for the entropy in Fourier space in
equation (\ref{entropy:vN:Delta_FourierTransform}) as:
\begin{equation}
S_{\rm vN} = V \int \frac{\mathrm{d}^{3}\mathbf{k}}{(2\pi)^3}
\left[\frac{\bar \Delta (\mathbf{k},t)+1}{2} \ln\left(\frac{\bar
\Delta(\mathbf{k},t)+1}{2}\right) - \frac{\bar
\Delta(\mathbf{k},t)-1}{2} \ln\left(\frac{\bar
\Delta(\mathbf{k},t)-1}{2}\right) \right] \,,
\label{entropy:vN:Delta_FourierTransformNGcase}
\end{equation}
Assuming that $\zeta_i(\mathbf{k},t) \ll 1$ we can derive the
change in entropy to linear order in $\zeta_i$. The result is:
\begin{subequations}
\label{s2:finalQFT}
\begin{equation}
S_2(t)=S_{\mathrm{g}}(t) + \delta S_{2}(t) \label{s2:finalQFTa}
\,,
\end{equation}
where $S_{\mathrm{g}}$ is the Gaussian contribution and $\delta
S_{2}$ is an entropy shift given by:
\begin{eqnarray}
\delta S_2 &=&  V \int \frac{\mathrm{d}^{3}\mathbf{k}}{(2\pi)^3}
\Bigg[\left(\frac{\zeta_{1R}(\mathbf{k},t)}{A_{\mathrm{R}}(\mathbf{k},t)-C(\mathbf{k},t)}+
\frac{1+2n(\mathbf{k},t)+2n^2(\mathbf{k},t)}{2n(\mathbf{k},t)(1+n(\mathbf{k},t))}\frac{\zeta_2(\mathbf{k},t)}{A_{\mathrm{R}}(\mathbf{k},t)-C(\mathbf{k},t)}\right)
\nonumber \\
&& \qquad\qquad\qquad \times
\frac{n(\mathbf{k},t)(1+n(\mathbf{k},t))}{1+2n(\mathbf{k},t)}
\ln\left(\frac{1+n(\mathbf{k},t)}{n(\mathbf{k},t)}\right)\Bigg]
\label{s2:finalbQFT}\,,
\end{eqnarray}
\end{subequations}
where, of course, $n(\mathbf{k},t) = (\Delta(\mathbf{k},t)-1)/2$.
This result is the field theoretic generalisation of the quantum
mechanical entropy~(\ref{s2:final}). It can be applied to mildly
non-Gaussian states in field theory, which, apart from being
mixed, also contain small one particle contributions.

The field theoretical generalisation of the second example
presented in section \ref{Kurtosis} is hard to solve for, even in
the homogeneous case, as it is non-local in Fourier space.

\subsection{Example: A Scalar Field with a Changing Mass}
\label{Example: A Scalar Field with a Changing Mass}

As a simple illustration of the ideas presented above, let us
investigate the effect of a changing mass on the Gaussian entropy
for a scalar field. This is maybe not a very exciting example, as
no entropy is generated of course in free theories, it
nevertheless provides an intuitive way of how the Wigner function
can be used. Let us consider the action of a free scalar field:
\begin{equation}\label{action1}
S[\phi] = \int \mathrm{d}^{4}\!x \left\{-\frac{1}{2}
\partial_\mu\phi(x)
\partial_\nu \phi(x) \eta^{\mu\nu} - \frac{1}{2} m^{2}_{\phi}(t)
\phi^{2}(x) \right\}\,,
\end{equation}
where as usual $\eta_{\mu\nu} = {\rm diag}(-1,1,1,1)$ is the
Minkowski metric, and where we consider the following behaviour of
the mass $m_{\phi}(t)$ of the scalar field, mediated for example
by some other Higgs-like scalar field:
\begin{equation}\label{masstanh}
m^{2}_{\phi}(t) = \left(A + B \tanh(\rho t)\right)\,.
\end{equation}
The equation of motion following from (\ref{action1}) reads:
\begin{equation} \label{eom1}
\left(\partial_{t}^{2}-\partial_{i}^{2}+m^{2}_{\phi}(t) \right)
\phi(x) =0\,,
\end{equation}
Let us quantise our fields in $D$-dimensions by making use of
creation and annihilation operators:
\begin{equation} \label{quantisationphi}
\hat{\phi}(x) = \int
\frac{\mathrm{d}^{\scriptscriptstyle{D}-1}\vec{k}}{(2\pi)^{\scriptscriptstyle{D}-1}}
\left( \hat{a}_{\vec{k}}\,\phi_{k}(t)e^{i\vec{k} \cdot \vec{x}} +
\hat{a}_{\vec{k}}^{\dag}\,\phi_{k}^{\ast}(t)e^{-i\vec{k} \cdot
\vec{x}} \right)\,.
\end{equation}
The annihilation operator acts on the vacuum as usual
$\hat{a}_{\vec{k}}|0\rangle = 0$. We impose the following
commutation relations: $[ \hat{a}_{\vec{k}},
\hat{a}_{\vec{k'}}^{\dag}] =(2\pi)^{\scriptscriptstyle{D}-1}
\delta^{\scriptscriptstyle{D}-1}(\vec{k}-\vec{k'})$. The mode
functions $\phi_{k}(t)$ of $\phi(x)$, defined by relation
(\ref{quantisationphi}) thus obey:
\begin{equation}\label{eom2}
\left(\partial_{t}^{2} + \omega^{2}(t) \right) \phi_{k}(t) = 0 \,,
\end{equation}
where $k=\|\vec{k}\|$ and $\omega^{2}(t)=k^{2} + m^{2}_{\phi}(t)$.
Using equation (\ref{F:1}) with
$\hat{\rho}(t_{0})=|0\rangle\langle0|$, we see that the mode
functions determine the statistical propagator completely:
\begin{equation}
F_{\phi}(k,t,t') = \frac{1}{2} \left\{
\phi_{k}(t')\phi_{k}^{\ast}(t) +
\phi_{k}(t)\phi_{k}^{\ast}(t')\right\}
\label{statisticalpropagatorF} \,.
\end{equation}
The solution of (\ref{eom2}) which behaves as a positive frequency
mode in the asymptotic past, i.e.: $\lim_{t\rightarrow - \infty}
\phi_{k}^{\mathrm{in}}(t) = \exp\left[- \imath
\omega_{\mathrm{in}} t \right]/\sqrt{2\omega_{\mathrm{in}}}$, can
be expressed in terms of Gauss' hypergeometric function
${}_{2}F_{1}$ (see \cite{Bernard:1977pq, Birrell:1982ix}):
\begin{equation} \label{modesolution1}
\phi_{k}^{\mathrm{in}}(t) = \frac{1}{\sqrt{2\omega_{\mathrm{in}}}}
\exp \left[- \imath \omega_{+} t - \imath
\frac{\omega_{-}}{\rho}\log\{2\cosh(\rho t )\}\right]
\phantom{1}_{2}F_{1}\left( 1+ \imath \frac{\omega_{-}}{\rho},
\imath \frac{\omega_{-}}{\rho}; 1- \imath
\frac{\omega_{\mathrm{in}}}{\rho};\frac{1}{2}\{1+\tanh(\rho
t)\}\right) \,,
\end{equation}
where we defined $\omega_{\mathrm{in}} = ( k^{2}+A-B
)^{\frac{1}{2}}$, $\omega_{\mathrm{out}} = (
k^{2}+A+B)^{\frac{1}{2}}$ and $\omega_{\pm}= (
\omega_{\mathrm{out}} \pm \omega_{\mathrm{in}})/2$. Having the
mode functions at our disposal, we can find the rather cumbersome
expressions for the exact statistical propagator. The statistical
propagator in turn fixes the phase space area through equation
(\ref{Delta function:phi:momentum}), yielding $\Delta_{k}(t) = 1$
such that:
\begin{equation}\label{solS2}
S_{k}(t) = 0 \,.
\end{equation}
\begin{figure}[t!]
    \begin{minipage}[t]{.43\textwidth}
        \begin{center}
\includegraphics[width=\textwidth]{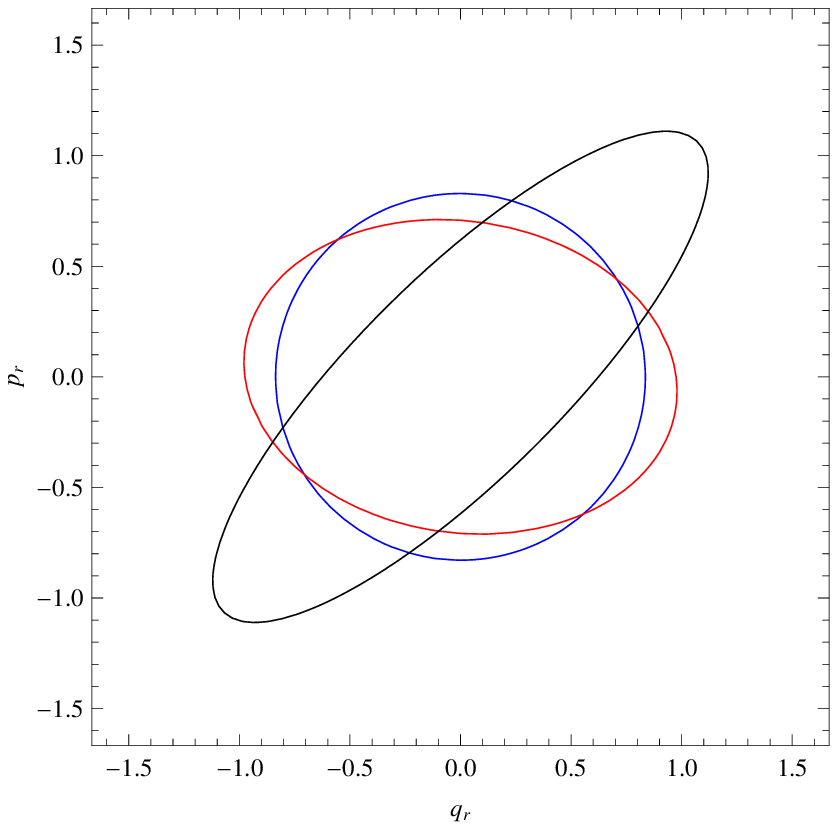}
   {\em \caption{Squeezing of the Gaussian phase space in the Wigner representation
   due to a changing mass. We
   used $k/\rho= 0.2$ and $m/\rho$ changes from 0 to 4 where the mass changes most rapidly at $t\rho=5$. For early
   times $t\rho=0$, the Gaussian phase space is a perfect circle (blue).
   Already for $t\rho=2$ a little squeezing is visible (red) and at late
   times $t\rho=9$ this is manifest (black). The coordinates $p_{r}$ and $q_{r}$
   have been rescaled for dimensional reasons:
   $p_{r}=p\sqrt{\omega(t)}$ and $q_{r}=q/\sqrt{\omega(t)}$.
    \label{fig:Wigner1}}}
        \end{center}
    \end{minipage}
\hfill
    \begin{minipage}[t]{.43\textwidth}
        \begin{center}
\includegraphics[width=\textwidth]{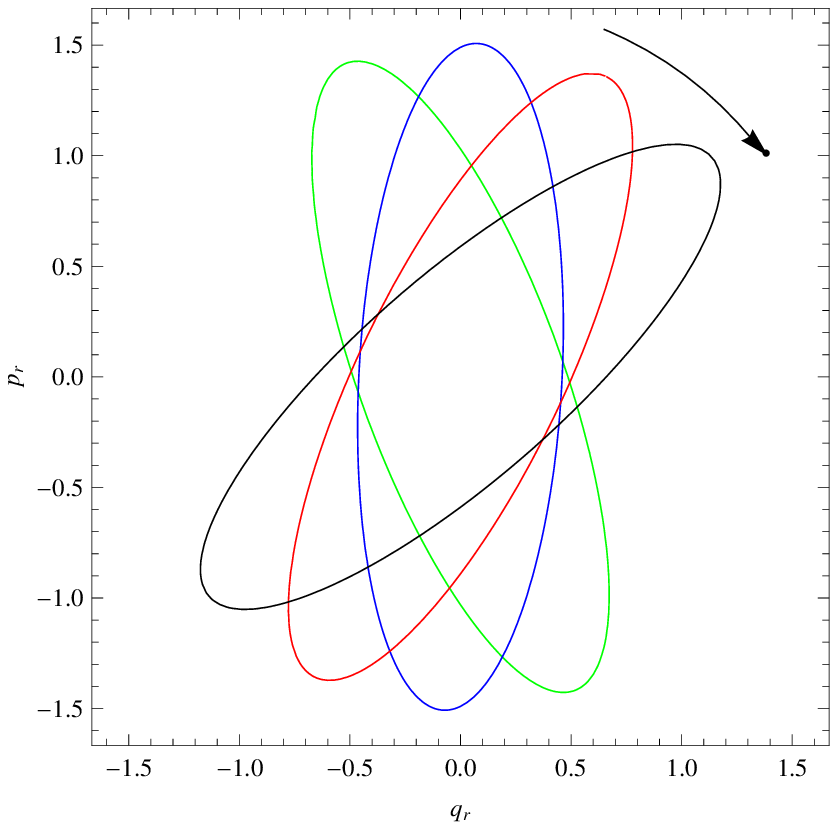}
   {\em \caption{Rotation of the Gaussian phase space in the Wigner representation
   due to a mass that has changed. We used the same parameters as
   in figure \ref{fig:Wigner1}. The squeezed phase space ellipses
   are plotted at times $t\rho=9.5$ (green), $t\rho=9.6$ (blue),
   $t\rho=9.7$ (red) and $t\rho=9.8$ (black). The rotation is clearly visible and its direction is indicated by the arrow. This can be
   understood from realising that an arbitrary squeezed state can
   be interpreted as a superposition of ordinary coherent states
   and the knowledge that the latter show the same rotating
   behaviour. \label{fig:Wigner2}}}
        \end{center}
    \end{minipage}
\end{figure}
We thus conclude that a changing mass does not change the entropy
for a free scalar field. Let us now examine the same process in
Wigner space. Of course, we will reach a similar conclusion
(\ref{solS2}) but Wigner space is much more suited to visualise
the process neatly. The statistical propagator is also the
essential building block for the Wigner function which can be
appreciated from generalising equations (\ref{aI:aR:c}) and
(\ref{Wigner function:Gauss2}) to:
\begin{subequations}
\label{Wignercoefficients}
\begin{eqnarray}
\alpha_{\mathrm{w}}(k,t) &=& \frac{1}{2 F_{\phi}(k,t,t)} \label{Wignercoefficientsa}\\
\beta_{\mathrm{w}}(k,t) &=& \frac{2
F_{\phi}(k,t,t)}{\Delta_{k}^{2}(t)}
\label{Wignercoefficientsb}\\
p_{\mathrm{c}} &=& - \frac{\partial_{t}
F_{\phi}(k,t,t')|_{t=t'}}{F_{\phi}(k,t,t)}
\label{Wignercoefficientsc} \\
\mathcal{M}_{k}(t) &=& \frac{2}{\Delta_{k}(t)}
\label{Wignercoefficientsd} \,.
\end{eqnarray}
\end{subequations}
We can thus plot cross-sections of the 2-dimensional Gaussian
phase space distribution in the Wigner representation. In figure
\ref{fig:Wigner1} we depict the squeezing of the Gaussian phase
space due to the changing mass. At early times, the state is still
in an unsqueezed vacuum. At $t\rho=2$ some squeezing is visible,
whereas at late times $t\rho=9$ this is manifest. Despite the
effect of the changing mass on the accessible phase space in
Wigner space, the area of the ellipse remains constant throughout
the whole process as expected from equation (\ref{solS2}). At late
times, when the mass has settled to its final constant value
$m/\rho=4$, the squeezed ellipse rotates in Wigner space, which we
depict in figure \ref{fig:Wigner2}, but it is not squeezed any
further. The rotation can be anticipated by the intuitive notion
that an arbitrary squeezed state can be thought of as a
superposition of coherent states (just imagine that the ellipse is
replaced by a number of circles displaced from the origin).
Coherent states that are displaced from the origin rotate in time.

\section{Conclusion}
\label{Conclusion}

We have formally developed a novel approach to decoherence in
quantum field theory: neglecting observationally inaccessible
correlators will give rise to an increase in the entropy of the
system. This is inspired by realising that correlators are
measured in quantum field theories and that higher order n-point
functions are usually perturbatively suppressed. An important
advantage of this approach is that the procedure of
renormalisation can systematically be implemented in this
framework.

We have shown how knowledge about the correlators of a system
affects the notion of the entropy associated to that state in two
cases: we firstly calculated the entropy of a general Gaussian
state. This entropy can be expressed purely in terms of the three
equal time correlators characterising the Gaussian state.
Moreover, all three correlators can be obtained from the
statistical propagator which opens the possibility to study
quantum corrections on the entropy in an interacting quantum field
theory in a systematic manner. Secondly, we calculated the entropy
for two specific types of non-Gaussian states. Firstly, we assumed
that the state of the system can be described by an admixture of
the ground state and a small contribution of the first excited
state. This yields a small correction to the Gaussian entropy.
Secondly, in the quantum mechanical case, we calculated an
expression for the entropy in case our observer could probe a
specific type of kurtosis of the ground state. This also changes
the entropy that the observes associates to the state.

We also outlined the use and limitations of the Wigner function,
the Wigner transform of the density matrix of a system. In
particular, we have shown that $\Delta$, the fundamental quantity
constructed from various correlators that fixes the entropy,
indeed coincides with the phase space area in Wigner space.

This is a rather phenomenological discussion, connecting the
notion of entropy, correlators and phase space area in quantum
field theory. Although we have outlined various mechanisms how
entropy could be generated, we have in the present paper not
applied our ideas to concrete systems. We refer the reader to e.g.
\cite{Giraud:2009tn, Koksma:2009wa, Koksma:2010} for specific
examples of how entropy is generated in interacting quantum field
theories in an out-of-equilibrium setting.

\

\noindent \textbf{Acknowledgements}

\noindent JFK and TP thank Theo Ruijgrok and gratefully
acknowledge the financial support from FOM grant 07PR2522 and by
Utrecht University. The authors also gratefully acknowledge the
hospitality of the Nordic Institute for Theoretical Physics
(NORDITA) during their stay at the ``Electroweak Phase
Transition'' workshop in June, 2009.

\appendix

\section{Useful Integrals}
\label{Appendix A Useful integrals}

Here we quote some integrals that are used in
section~\ref{Non-Gaussian entropy: Two examples}. The three
integrals are:
\begin{eqnarray}
 I_{\eta 0}(\epsilon)&=&\int_{-\infty}^\infty \mathrm{d}x_0 \cdots \int_{-\infty}^\infty \mathrm{d}x_\epsilon
 \left(\eta_0 x_\epsilon^4 + \eta_0^* x_0^4 \right)
  \exp \left[-2a_{\mathrm{R}}(x_0^2 + \cdots + x_\epsilon^2)+2c(x_0x_1+\cdots + x_\epsilon x_0) \right]
 \nonumber\\
  &&\hskip 5cm
  =\, (\eta_0 + \eta_0^*)\frac34\frac{\pi^{\frac{1+\epsilon}{2}}}
                                     {(2\beta)^\frac{5+\epsilon}{2}}
  \frac{(1-\alpha^{2+2\epsilon})^2}{(1-\alpha^2)^2(1-\alpha^{1+\epsilon})^5}
\label{App:eta0:int}
\\
I_{\eta 1}(\epsilon)&=&\int_{-\infty}^\infty \mathrm{d}x_0 \cdots
\int_{-\infty}^\infty \mathrm{d}x_\epsilon
 \left (\eta_1 x_\epsilon^3x_0 + \eta_1^* x_\epsilon x_0^3 \right)
  \exp \left[-2a_{\mathrm{R}}(x_0^2 + \cdots + x_\epsilon^2)+2c(x_0x_1+\cdots + x_\epsilon x_0) \right]
 \nonumber\\
  &&\hskip 5cm
  =\,  (\eta_1 + \eta_1^*)\frac34\frac{\pi^{\frac{1+\epsilon}{2}}}
                                     {(2\beta)^\frac{5+\epsilon}{2}}
  \frac{\alpha(1+\alpha^{-1+\epsilon})(1-\alpha^{2+2\epsilon})}
       {(1-\alpha^2)^2(1-\alpha^{1+\epsilon})^4}
\label{App:eta1:int}
\\
I_{\eta 2}(\epsilon)&=&\int_{-\infty}^\infty \mathrm{d}x_0 \cdots
\int_{-\infty}^\infty \mathrm{d}x_\epsilon
 \left(\eta_2 x_\epsilon^2 x_0^2 \right)
  \exp \left[-2a_{\mathrm{R}}(x_0^2 + \cdots + x_\epsilon^2)+2c(x_0x_1+\cdots + x_\epsilon x_0) \right]
 \nonumber\\
  &&\hskip 5cm
  =\,  \eta_2\frac14\frac{\pi^{\frac{1+\epsilon}{2}}}
                                     {(2\beta)^\frac{5+\epsilon}{2}}
  \frac{(1+\alpha^2)^2(1+\alpha^{-2+2\epsilon})
            -\alpha^4(1+\alpha^{-6+2\epsilon})+6\alpha^{1+\epsilon}}
       {(1-\alpha^2)^2(1-\alpha^{1+\epsilon})^3}
       \,,\qquad
\label{App:eta2:int}
\end{eqnarray}
where $\epsilon\geq 0$ is an integer.
 The first non-trivial check of these integrals is to set $\epsilon=0$, in which case their sum
 gives the (inverse) normalisation constant~(\ref{calN4}),
\begin{equation}
  I_{\eta 0}(0)+ I_{\eta 1}(0) +  I_{\eta 2}(0) = \sqrt{\frac{\pi}{2(a_{\mathrm{R}}-c)}}\frac{3}{4}
   \frac{\eta_0+\eta_0^*+\eta_1+\eta_1^*+\eta_2}{[2(a_{\mathrm{R}}-c)]^2}
    = {\cal N}_4^{-1}-{\cal N}_{\rm g}^{-1}
\,. \nonumber
\end{equation}
Moreover, we encourage the reader to check the integrals for some
other values of $\epsilon$. By performing direct integrations we
have checked the
integrals~(\ref{App:eta0:int}--\ref{App:eta2:int}) for $\epsilon =
1,2,3,4,5$, and they all give the correct answer. Since the
expressions~(\ref{App:eta0:int}--\ref{App:eta2:int}) are analytic
in $\epsilon$ when $0<\alpha<1$, they can be uniquely analytically
extended to complex $\epsilon$'s in that range of $\alpha$, based
on which we calculate the von Neumann entropy in
section~\ref{Non-Gaussian entropy: Two examples}.
 For the entropy calculation we need to expand
the integrals~(\ref{App:eta0:int}--\ref{App:eta2:int}) around
$\epsilon=0$ up to linear order in $\epsilon$. From
equation~(\ref{rho4:lin}) we see that it is convenient to multiply
the integrals
 with ${\cal N}_{\rm g}^{1+\epsilon}= [2(a_{\mathrm{R}}-c)/\pi]^{(1+\epsilon)/2}$,
\begin{eqnarray}
{\cal N}_{\rm g}^{1+\epsilon}I_{\eta 0}(\epsilon) &=&\frac34
\frac{\eta_0 + \eta_0^*}{[2(a_{\mathrm{R}}-c)]^2}
 \bigg\{
      1+\epsilon\bigg[
                  \ln(1-\alpha) + \frac{\alpha(5+\alpha)}{1-\alpha^2}\ln(\alpha)
               \bigg]
 \bigg\}
\label{App:eta0:int:b}
\\
{\cal N}_{\rm g}^{1+\epsilon}I_{\eta 1}(\epsilon)
 &=&
   \frac34\frac{\eta_1 + \eta_1^*}{[2(a_{\mathrm{R}}-c)]^2}
 \bigg\{
      1+\epsilon\bigg[
                  \ln(1-\alpha) +\frac{1+3\alpha+2\alpha^2}{1-\alpha^2}\ln(\alpha)
               \bigg]
 \bigg\}
\label{App:eta1:int:b}
\\
{\cal N}_{\rm g}^{1+\epsilon}I_{\eta 2}(\epsilon)
 &=&
   \frac34\frac{\eta_2}{[2(a_{\mathrm{R}}-c)]^2}
 \bigg\{
      1+\epsilon\bigg[
                  \ln(1-\alpha)
                  + \frac13\frac{4+7\alpha+7\alpha^2}{1-\alpha^2}\ln(\alpha)
               \bigg]
 \bigg\}
       \,,\qquad
\label{App:eta2:int:b}
\end{eqnarray}
where we used $a_{\mathrm{R}}/c = (1+\alpha^2)/(2\alpha)$, $2\beta
= c/\alpha = 2(a_{\mathrm{R}}-c)/(1-\alpha)^2$, and we rescaled
$\eta_i$ by the Gaussian width of the state squared,
$(a_{\mathrm{R}}-c)^2$ to write it in natural dimensionless units.

\section{Von Neumann Entropy for a Gaussian Field Theory}
\label{Appendix B Entropy for Gaussian Field Theory}

Here we outline how to calculate the von Neumann entropy in field
theory for a quantum state that can be described by a general
Gaussian density matrix. Keeping equation (\ref{replica trick}) in
mind, we realise that, as in the quantum mechanical case, we have
to evaluate:
\begin{eqnarray}
{\rm Tr}\left[\hat\rho^{1+\epsilon}\right] &=& {\cal
N}^{1+\epsilon}\! \int\! {\cal D}\phi_0 \cdots\! \int {\cal
D}\phi_{\epsilon} \exp\Big[-2(\phi_0^T\!\cdot
A_{\mathrm{R}}\cdot\phi_0 + \cdots + \phi_\epsilon^T\!\cdot
A_{\mathrm{R}}\cdot\phi_\epsilon)
\label{replica trick:field:1} \\
&&\qquad\qquad\qquad\qquad\qquad\qquad\qquad + \phi_0^T\!\cdot
C\cdot\phi_1 + \phi_1^T\!\cdot C\cdot\phi_0  + \cdots +
\phi_\epsilon^T\!\cdot C\cdot\phi_0 + \phi_0^T\!\cdot
C\cdot\phi_\epsilon \Big]
\nonumber\\
&=& {\cal N}^{1+\epsilon}\! \int\! {\cal D} \breve\phi_0 \cdots\!
\int {\cal D} \breve\phi_{\epsilon}
\exp\Big[-2(\breve\phi_0^T\!\cdot \breve{A}_{\mathrm{R}}
\mathbb{I} \cdot \breve\phi_0 + \cdots +
\breve\phi_\epsilon^T\!\cdot \breve{A}_{\mathrm{R}} \mathbb{I}
\cdot \breve\phi_\epsilon)
\nonumber \\
&&\qquad\qquad\qquad\qquad\qquad\qquad\qquad +
\breve\phi_0^T\!\cdot \breve{C} \mathbb{I} \cdot \breve \phi_1 +
\breve \phi_1^T\!\cdot \breve C \mathbb{I} \cdot \breve \phi_0  +
\cdots + \breve\phi_\epsilon^T\!\cdot \breve{C} \mathbb{I} \cdot
\breve \phi_0 + \breve \phi_0^T\!\cdot \breve{C} \mathbb{I} \cdot
\breve\phi_\epsilon \Big]
\nonumber\\
&=& {\cal N}^{1+\epsilon}J\! \int\! {\cal D}\tilde\phi_0 \cdots\!
\int {\cal D}\tilde\phi_{\epsilon}
\exp\left[-2(\tilde\phi_0^T\!\cdot \beta \mathbb{I }\cdot\tilde
\phi_0 + \cdots + \tilde\phi_\epsilon^T\!\cdot \beta \mathbb{I}
\cdot\tilde\phi_\epsilon)\right] \nonumber \,,
\end{eqnarray}
where we note that $A_{\mathrm{R}}$ and $C$ are real and symmetric
matrices and we assume that they can both be diagonalised by the
same orthogonal matrix $O$. We transformed to the diagonal field
coordinates by setting e.g. $\breve\phi_0=O \phi_0$, such that all
matrices diagonalise, e.g.: $O^{-1}\cdot A_{\mathrm{R}} \cdot O =
\breve{A}_{\mathrm{R}}(\vec x,t) \mathbb{I}$. Moreover, we assumed
that $C^T=C$. Secondly, we transformed variables to $ \tilde\phi_i
= \breve\phi_i-\alpha \mathbb{I} \breve\phi_{i+1}$ for
$i=\{0,1,\cdots,\epsilon-1\}$ and $\tilde\phi_\epsilon =
\breve\phi_\epsilon-\alpha \mathbb{I} \breve\phi_0$, analogous to
the quantum mechanical case. The Jacobian of this change of
variables is now given by:
\begin{equation}
J =  \left|\frac{\partial(\breve \phi_i)}{\partial(\tilde
\phi_j)}\right| = \left| {\rm det}[R^{-1}]\right| = {\rm det}
\left [\left(\mathbb{I}-\alpha \mathbb{I} \right)^{-1} \right]\,,
\end{equation}
where we write down the matrix $R$ for clarity as:
\begin{equation}
R =  \left(\begin{array}{ccccc}
\mathbb{I} & -\alpha \mathbb{I} & 0 & \cdots & 0 \\
0 & \mathbb{I} & -\alpha \mathbb{I} & \cdots &  0    \\
0 &  0& \mathbb{I} & \cdots &   0     \\
\vdots & \vdots & \phantom{1} & \ddots & \vdots \\
-\alpha \mathbb{I}  & 0  & \cdots & 0 & \mathbb{I} \\
\end{array}\right)
\,. \label{Rd:phi}
\end{equation}
The diagonal matrices $\alpha(\vec x,t)\mathbb{I}$ and $\beta(\vec
x,t)\mathbb{I}$ generalise to:
\begin{eqnarray}
\alpha(\vec x,t)\mathbb{I} &=& \left( \frac{\breve
A_{\mathrm{R}}(\vec x,t)}{\breve C(\vec x,t)}
-\sqrt{\left(\frac{\breve A_{\mathrm{R}}(\vec x,t)}{\breve C(\vec
x,t)}\right)^2-1}
\right)\mathbb{I} \label{replica trick:alpha-beta:phi}\\
\beta(\vec x,t)\mathbb{I} &=& \frac{\breve C(\vec x,t)}{2
\alpha(\vec x,t)}\mathbb{I} \,,
\end{eqnarray}
where we note that $\breve{A}_{\mathrm{R}} - \breve{C} = \beta
(1-\alpha)^2$. Equation~(\ref{replica trick:field:1}) can now
easily be evaluated:
\begin{equation}
{\rm Tr}\left[\hat\rho^{1+\epsilon}\right] = {\rm det}\left[\left
(\frac{(1-\alpha)^{\epsilon+1}}{1-\alpha^{\epsilon+1}}\right)\mathbb{I}
\right] \,. \label{replica trick:field:3}
\end{equation}
Expanding around $\epsilon=0$ we get:
\begin{equation}
S_{\rm vN} = -{\rm Tr}\left[ \left(
\ln(1-\alpha)+\frac{\alpha}{1-\alpha} \ln(\alpha)\right)\mathbb{I}
\right] \,, \label{entropy:vN:alpha:d}
\end{equation}
where we made use of, ${\rm det}[\mathbb{I}+\epsilon A] = 1 +
\epsilon{\rm Tr}[A]+{\cal O}(\epsilon^2)$. The phase space
invariant $\Delta^{2}(\vec x, \vec y,t)$ in equation (\ref{Delta
function:phi}) can also be cast in diagonal form:
\begin{equation}
\breve\Delta^{2}(\vec x,t) \mathbb{I} =
\frac{\breve{A}_{\mathrm{R}}(\vec x,t)+\breve{C}(\vec
x,t)}{\breve{A}_{\mathrm{R}}(\vec x,t)-\breve{C}(\vec x,t)}
\mathbb{I} \label{diagonalDelta}
\end{equation}
We can thus derive $\alpha(\vec x,t)
=(\breve\Delta(\vec{x},t)-1)/(\breve\Delta(\vec x,t)+1)$. The
resulting equation for the entropy can be rotated back to the
original non-diagonal field coordinates by making use of the
unitary matrix $O$:
\begin{equation}
S_{\rm vN}
 = {\rm Tr}\left[\frac{\Delta+\mathbb{I}}{2}\cdot\ln\bigg(\frac{\Delta+\mathbb{I}}{2}\bigg)
                - \frac{\Delta-\mathbb{I}}{2}\cdot\ln\bigg(\frac{\Delta-\mathbb{I}}{2}\bigg)
                   \right]
\,, \label{entropy:vN:Delta2}
\end{equation}
which is nothing but equation (\ref{entropy:vN:Delta}).

\end{document}